\documentstyle[11pt,epsfig]{article}
\newcommand{\be}{\begin{equation}}
\newcommand{\ee}{\end{equation}}
\newcommand{\beqn}{\begin{eqnarray}}
\newcommand{\eeqn}{\end{eqnarray}}

\newcommand{\frho}{{\bf \rho}}
\newcommand{\qf}{{\bf q}}
\newcommand{\kf}{{\bf k}}
\newcommand{\lf}{{\bf l}}
\newcommand{\mf}{{\bf m}}

\newcounter{savefig}
\newcommand{\alphfig}{\setcounter{savefig}{\value{figure}}%
\setcounter{figure}{0}%
\renewcommand{\thefigure}{\mbox{\arabic{savefig}\alph{figure}}}}
\newcommand{\resetfig}{\setcounter{figure}{\value{savefig}}%
\renewcommand{\thefigure}{\arabic{figure}}}
\begin{document}
\baselineskip=24pt
\begin{center}
\begin{large}
The BFKL Pomeron in Deep Inelastic Diffractive Dissociation near
$t=0$
\vspace{1cm}

J. Bartels, H.Lotter and M.W\"usthoff \\
\end{large}
{\it II. Institut f\" ur Theoretische Physik, Universit\" at Hamburg}
\\
\end{center}

\vspace{5cm}

{\bf Abstract:}
The small-$t$ behaviour of the deep inelastic diffractive dissociation
cross section in the triple Regge region is investigated, using
the BFKL approximation in perturbative QCD.
We show that the cross section is finite at $t=0$, but the diffusion
in $\ln{k_t^2}$ leads to a large contribution of small momenta at the
triple Pomeron vertex. We study the dependence upon the total energy
and the invariant mass. At $t=0$, there is a decoupling of the three
BFKL singularities which is a consequence of the conservation of the
conformal dimension. For large invariant masses, the four gluon state
in the upper t-channel plays an important role and cannot be neglected.
\newpage
\section{Introduction}
\setcounter{equation}{0}
The study of perturbative QCD in the triple-Regge limit has recently
attracted some interest ~\cite{MueP,BW}. If one interpretes the
observed strong rise of $F_2$ at small x as a signal for the
BFKL-Pomeron ~\cite{BFKL}, it is natural to ask for corrections to
this new piece of perturbative QCD, and an obvious place to look
for such terms is the triple Regge limit. There may also be some
interest in this limit from an experimental point of view: some of the
observed ``rapidity gap events`` ~\cite{Zeus,H1} may belong to a
kinematical region where perturbative QCD is applicable.

In ~\cite{BW} an attempt has been made to derive an analytic
formula for the triple Regge inclusive cross section which lies at the
same level of accuracy as the BFKL Pomeron. The result was given
in a somewhat abstract form, and, so far, only a few rather
general properties have been studied. The general structure of
the cross section formula is illustrated in Fig.1a: starting from the
top, the fermion box first couples to a BFKL-ladder. At the transition:
two gluons $\rightarrow$ four gluons a new vertex function appears.
Below this vertex the four-gluon
state starts where the gluons interact pairwise in all possible ways.
Finally, the four-gluon state branches into the two BFKL ladders
at the bottom. In addition to this general structure, there is also
a contribution where the upper BFKL-ladder couples directly to the
lower ones (Fig.1b).

Based upon the experience with BFKL Pomeron, we expect that
the formalism developed in \cite{BW} is suited to study the whole range
of momentum transfer $t$ of the diffractive dissociation cross section
(provided that $\sqrt{-t}$ is still smaller than $M$, the missing
mass of produced hadronic system). Nevertheless, there are several
reasons to believe that the point $t=0$ plays a very special role
and perturbation theory may even not be applicable at all. Firstly,
the early hard scattering approach of ref. \cite{GLR,Rys}
(and similar calculations later on) has shown that the large transverse
momenta at the two gluons $\rightarrow$ four gluons- vertex (fig. 1a)
are suppressed like $dk_t^2/k_t^4$, i.e. small transverse momenta
dominate. Without invoking further corrections, the cross section
would diverge at $k_t^2=0$: in the framework of the GLR equation
which leads to a saturation of the Pomeron
it is the unitarity corrections (screening) to the lower Pomerons
which provide the necessary supression at small $k_t^2$. Since the
saturation begins at a rather large momentm scale
($2 - 4\, GeV^2$, depending upon the ratios $M^2/s$ ~\cite{Rys}),
this mechanism tends to predict hard final states. Secondly,
also within BFKL physics
the point $t=0$ is exceptional: whithin the lower Pomerons
the diffusion in $\ln{k_t^2}$ extends into both the ultraviolet and the
infrared regions, whereas for $t \neq 0$ the diffusion into the
infrared region where perturbation theory becomes unreliable
is stopped by the momentum scale $t$. One therefore
expects, for the point $t=0$, the ``dangerous`` region of small $k_t$
to play a much more important role than for the case $t\neq 0$.
This expectation has recently been confirmed by
Mueller ~\cite{MueP}, using the large-$N_c$ approximation. As a result
of this approximation, the four gluon state above the triple
Pomeron vertex is absent, and one is lead directly to a study of the
diagrams shown in Fig.1b. The final result of this study is an explicit
formula for the inclusive cross section, showing the dependence
upon the energy variables $s$, $M$, and the momentum transfer $t$.
The latter one is of particular interest: for an intermediate
$t$-region, the cross section goes as $1/\sqrt{-t}$. This behaviour
hints at some sort of singular behaviour at $t=0$, in agreement with
what one might expect in the diffusion picture. From this study,
however, it is not clear what happens at $t=0$, in particular, whether
the whole perturbative analysis breaks down or not.

The more general reason why, from the theoretical point of view, it is
important to understand the small-t behaviour of the diffractive
dissociation cross section is the problem of unitarization. It is
well-known that the sum of the leading logarithms
at sufficiently large energies (or sufficiently small $x_{Bj}$
runs into conflict with unitarity, and the way in which unitarity is
restored is still an open question. Diffractive dissociation is not
contained in the leading logarithmic approximation and, therefore,
represents a (observable) correction which contributes to the
unitarization procedure.
Within the GLR-equation, the finite limit of the cross section
at $t=0$ requires the complete unitarization (saturation) of the lower
Pomerons.
Similarly, in ~\cite{Mue} it was shown that, while single unitarity
corrections have large infrared contributions, their resummation
leads to strong cancellations. In both cases, one expects the final
state to be rather hard. The process of diffractive
dissociation may therefore become a very sensitive tool in exploring
the unitarization mechanism. For example, one might count the number of
events in dependence on a lower cutoff on $k_t$ of the final state
or, alternatively, measure the momentum transfer $t$. At which scale
the saturation
occurs is still an open question, since the GLR-equation is only a
crude estimation compared to a complete procedure of unitarization. The
HERA data will soon show, whether a value for $k_t^2$ of about $2 - 4\,
GeV^2$ is justified.

In this paper, we are going to investigate the small t-region of the
cross section formula of ~\cite{BW}, in particular the point $t=0$.
As one of the main results of this paper we will show that the BFKL
approach to the diffractive dissociation, in spite of the
$dk_t^2/k_t^4$-behaviour at large $k_t$, is infrared safe,
and the limit $t=0$ exists. At the same time,
however, the BFKL diffusion has entered into the infrared region and
thus emphasizes the need to include unitarizing corrections to the
leading logarithmic approximation. We believe that this observation
is important from the point of view of theoretical consistency:
it shows that, within the BFKL approach, a smooth transition from a
finite $t$ down to $t=0$ is possible without running into infrared
singularities. Consequently, the approximation used in ~\cite{BW}
represents a well-defined starting point for approaching the
unitarization problem. Whether the (leading logarithmic) formula of
{}~\cite{BW} can
already be used for deducing experimental signatures remains less
clear. It is encouraging that a rough estimate of the ratio of
diffractive events over
all DIS-events gives a reasonable value. Furthermore, HERA-data, so
far, seem to support the dominance \cite{ZEUS} of events with low
$k_t$: this is in qualitative agreement with the strong diffusion
into the infrared region.
Therefore, it seems worthwile to study the dynamics of the
unscreened BFKL-Pomeron in more detail, even before adressing the
question of unitarization. Certain characteristics may very well
survive the unitarization procedure, and it is important to check
whether they may serve as signals to support or rule out the
BFKL-dynamics.

Apart from the result that the BFKL cross section formula has a
finite limit at $t=0$ our analysis contains a detailed saddle point
analysis and investigates, as a function of $t$ near $t=0$, the
dependence upon $s$,
$M^2$, and $Q^2$. For $|t|$ of the order of $Q^2$, our analysis confirms
the $1/\sqrt{-t}$ behaviour found in ~\cite{Mue} in the large-$N_c$-
approximation. Moving towards smaller $t$-values, the shape of the
t-distribution changes, and the cross section reaches, at $t=0$, a
finite limit. At the same time, the derivative with respect to $t$
tends to infinity.
For small $t$ one observes a shrinkage, i.e. the cusp becomes narrower
as $s\rightarrow \infty$ (the typical width shrinks with some inverse
power of $s/M^2$).
All these changes as a function of $t$ are accompanied by a very
peculiar dependence
upon $s$ and $M^2$. In particular, at $t=0$ one observes a decoupling
of the lower BFKL Pomerons from the BFKL singularity above.
Much of this striking behaviour can be traced back to the conformal
invariance of the BFKL approximation: at $t=0$ we find a conservation
law of the conformal dimension of the BFKL ladders above
and below the triple Pomeron vertex.

Our paper will be organized as follows. We begin with the simplest
case, the diffractive production of a $q \bar{q}$-pair in the
triple Regge region near $t=0$. This simple case already shows the
main result, namely the conservation law of conformal dimensions
and its implication for the high energy behavior. The advantage of
first presenting this simpler case lies if the fact that we are
able to present an analytic expression for the $M^2$-integrated
cross section which can directly be used for a comparison with
observed event rates. The discussion will first be done in
momentum space; in the subsequent section we repeat the derivation
in coordinate space where a more intuitive picture has been
developed \cite{Mue,NZ}. The generalization to the production of
$q \bar{q} + \mbox{gluons}$ will
be described in Section 4; since the analysis presented in this
part will be rather technical, we shall give a short summary at the
end of this section. In the final
section we discuss a few implications of the results of this paper.
\section{Diffractive Production of $q \bar{q}$-Pairs Near $t=0$}
\setcounter{equation}{0}
We begin with the $M^2$-integrated cross section for the process
(Fig.2a)
$\gamma^{\ast} + proton \rightarrow (q \bar{q}) + \mbox{proton}$,
where $M$
is the invariant mass of the quark pair, $t$ the square of the momentum
transferred from the proton to the quark pair, and $1/x_B = s/Q^2$
the total energy. We are interested in the limit of small $x_B$
and keep $t$ as a small variable parameter.

Following the notation of ~\cite{BW} we use the integral representation
\beqn
\frac{d \sigma}{dt} = \sum_{f} e_f^2
          \frac{\alpha_{em} (2 \pi)^3}{8 \pi Q^4}
 \int \frac{d \omega_1}{2 \pi i} \int \frac{d \omega_2}{2 \pi i}
 \left( \frac{1}{x_B} \right) ^{\omega_1 + \omega_2}
          F(\omega_1, \omega_2, t),
\label{dcs}
\eeqn
where the partial wave consists of the three building blocks
illustrated in Fig.2b :
at the upper end we have the four-gluon amplitude
$D_{(4;0)}^{(1;++)}$, below the two BFKL Pomerons, and at the bottom
we use a form factor for the coupling of the BFKL ladders to the
proton. The four-gluon amplitude has been studied in ~\cite{BW}.
It can be rewritten as a sum of two-gluon amplitudes:
\beqn
&&D_{(4,0)}^{(1;+,+)}(\kf_1,\kf_2,\kf_3,\kf_4)\;=\;g^2\;
\frac{\sqrt{2}}{3}\;\cdot\nonumber\\
&&\left\{\;D_{(2;0)}(\kf_1,\kf_2+\kf_3+\kf_4)\,
+\,D_{(2;0)}(\kf_2,\kf_1+\kf_3+\kf_4)\right.\\
&&+\,D_{(2;0)}(\kf_3,\kf_1+\kf_2+\kf_4)\,+\,
                 D_{(2;0)}(\kf_4,\kf_1+\kf_2+\kf_3)
                           \nonumber\\
&&\left.-\,D_{(2;0)}(\kf_1+\kf_2,\kf_3+\kf_4)\,-
                 \,D_{(2;0)}(\kf_1+\kf_3,\kf_2+\kf_4)\,
-\,D_{(2;0)}(\kf_1+\kf_4,\kf_2+\kf_3)\right\}\,\,.\nonumber
\label{d01}
\eeqn
For $D_{(2;0)} (\kf, -\kf)$ it is convenient to use a
Mellin transform with respect to the variable $k^2 / Q^2$:
\beqn
D_{(2;0)} (k^2) = \int \frac{d \mu}{2 \pi i} \left( \frac{k^2}{Q^2}
        \right) ^{- \mu} \tilde{D}_{(2;0)} (\mu),
\label{d02}
\eeqn
where the $\mu$-contour runs along the imaginary axis, intersecting
the real axis within the interval $(-1,0)$ (in the following we shall
use, as the intersection with the real axis, the point $-1/2$;
we shall then use the notation $\mu = -1/2 - i \nu$).
The function $\tilde{D}_{(2;0)} (\mu)$ has poles at positive and
negative integers, and a detailed discussion is contained in
{}~\cite{BW}. In this paper we only need the behaviour near $\mu  = -1$
\beqn
\tilde{D}_{(2;0)} \approx \sum_f \frac{e_f^2 \alpha_s \sqrt{8} }
                      {2 \pi} \frac{4}{3} \frac{1}{(\mu+1)^2}
\label{d20}
\eeqn
and near $\mu = -\frac{1}{2}$:
\beqn
\tilde{D}_{(2;0)} \approx \sum_f \frac{e_f^2 \alpha_s 9\sqrt{2} \pi^2}
                    {16}
\eeqn

Next we turn to the BFKL Pomeron. Since we want to study the $t$-
dependence for $t \neq 0$ we need an expression for the BFKL-pomeron
for non-zero momentum transfer.
The BFKL-pomeron is determined by a Bethe-Salpeter type of equation in
two dimensional transverse space. Lipatov \cite{Lip}
has shown that the configuration space representation of this equation
is invariant under two dimensional conformal transformations. Due to
this symmetry it can be diagonalized by a conformal partial wave
expansion. Using orthonormality and completeness of the
conformal partial waves Lipatov found an analytic expression for the
sum of the nonforward ladders. By a straightforward Fourier
transfomation, this expression leads to the following momentum
representation
\footnote{Our normalization differs from Lipatov's one by factors
of $2 \pi$ which are included in the integration measure in momentum
space}:
\beqn
\Phi_{\omega}(\kf,\kf',\qf) =
\int_{-\infty}^{+\infty} \frac{d \nu}{2 \pi}
          \frac{1}{\omega-\chi(0,\nu)}
   E^{(\nu)}(\kf,\qf-\kf) E^{(\nu) \ast} (\kf',\qf -\kf')
\label{pom}
\eeqn
where we have restricted ourselves to zero conformal spin.
The eigenvalues $\chi(0,\nu)$ of the BFKL-kernel are given by:
\beqn
\chi(0,\nu)  =  \frac{g^2 N_c}{4 \pi^2} \; [2\psi(1)-
         \psi(\frac{1}{2}+i \nu)-\psi(\frac{1}{2}-i \nu)]
\eeqn
The conformal partial waves have the momentum representation
\beqn
E^{(\nu)}(\kf,\qf-\kf) &=
                       & \frac{4^{-i\nu}}{4\pi}\;
\frac{\Gamma(1+2i\nu)}{\Gamma(-2i\nu)}
\frac{\Gamma(-\frac{1}{2}-i\nu)}{\Gamma(\frac{3}{2}+i\nu)}
\frac{\Gamma^2(\frac{1}{2}-i\nu)}{\Gamma^2(\frac{1}{2}+i\nu)}\;
        \int d^2 \rho_{1} d^2 \rho_{2}
        e^{i \kf \frho_1 + i (\qf - \kf)\frho_2}
     (\frac{\rho_{12}^2}{\rho_1^2\rho_2^2})^{\frac{1}{2} -i \nu}
         \nonumber \\
     & = & 2\pi\;
   \frac{\Gamma(1+2 i \nu)}{\Gamma(-2 i \nu)}
   \frac{\Gamma(-\frac{1}{2}-i \nu)}{\Gamma(\frac{1}{2}+i\nu)}
   \frac{\Gamma(\frac{3}{2}-i \nu)}{\Gamma(\frac{1}{2}+i\nu)}
           \nonumber \\ & & \cdot
           \int_0^1 d x [x(1-x)]^{-\frac{1}{2} +i \nu}
[\qf^2 x(1-x)+(\kf-x\qf)^2]^{-\frac{3}{2} -i \nu}
                   \nonumber \\
      & & \cdot _2F_1(\frac{3}{2}+i\nu,i \nu -
\frac{1}{2},1;\frac{(\kf-x\qf)^2}{\qf^2 x(1-x)+(\kf-x\qf)^2})
\label{pwm}
\eeqn
The normalization was chosen in such a way
\footnote{It differs from Lipatov's functions by some $\Gamma$ factors.}
that in the limit $\qf=0$ the expression (2.6) coincides with the
familiar BFKL-Pomeron in the forward direction:
\beqn
\Phi_{\omega}(\kf,\kf',\qf=0) =
      2(2\pi)^2 \int_{-\infty}^{+\infty}
          \frac{d \nu}{2 \pi}\frac{1}{\omega-\chi(0,\nu)}
(\kf^2)^{-\frac{3}{2}-i \nu} (\kf'^2)^{-\frac{3}{2}+i \nu}
\eeqn
In (2.8) one has to be careful
in taking the limit $q\rightarrow 0$, namely making use of the well known
properties of the hypergeometric functions one finds
\be
E^{(\nu)}(\kf,\qf \rightarrow 0)\;=\;2\pi [(k^2)^{-3/2-i\nu}\;+\;
C(\nu)(k^2)^{-3/2+i\nu}(q^2)^{-2i\nu}]
\ee
where
$C(\nu)$ is analytic in the range $-1/2<Im(\nu)<1/2$ and has the property
$C(-\nu)=1/C(\nu)$.
Hence, $E^{(\nu)}$ remains finite at $\qf =0$ only if $Im(\nu)>0$. If
$Im(\nu)<0$, $E^{(\nu)}$ becomes infinte. For $E^{(\nu)*}$ the converse is
true. So, one of the two factors, $E^{(\nu)}$ or $E^{(\nu)*}$, becomes
infinte, no matter what value of $\nu$ we choose. However, in (2.6) only
the product of the two $E^{(\nu)}$-functions appears:
$E^{(\nu)}(\kf,\qf \rightarrow 0)E^{(\nu)*}(\kf',\qf \rightarrow 0)=
(2\pi)^2[(\kf^2)^{-3/2-i\nu}(\kf'^2)^{-3/2+i\nu}+
(\kf^2)^{-3/2+i\nu}(\kf'^2)^{-3/2-i\nu}+
C(\nu)^2(\kf^2)^{-3/2-i\nu}(\kf'^2)^{-3/2-i\nu}(q^2)^{2i\nu}+
1/C(\nu)^2(\kf^2)^{-3/2+i\nu}(\kf'^2)^{-3/2+i\nu}(q^2)^{-2i\nu}$.
In the third and the fourth term, using the variable $\mu=-1/2-i\nu$,
 we have to shift the contour of integration  to the left and to the right,
respectively. Since $C(\nu)$ is analytic, both terms vanish. In the second
term we change from $\nu$ to $-\nu$. As a result we have
$\int d\nu\;E^{(\nu)}(\kf,\qf \rightarrow 0)E^
{(\nu)*}(\kf',\qf \rightarrow 0)=
 2\,(2\pi)^2\;\int d\nu\;(\kf^2)^{-3/2-i\nu}(\kf'^2)^{-3/2+i\nu}$, i. e.
we get the right answer, if we simply write $E^{(\nu)}(\kf,\qf\rightarrow 0)
=\sqrt{2}\,2\pi\;(k^2)^{-3/2-i\nu}$. In the following,
whenever we take the limit
$\qf \rightarrow 0$, we shall use this effective prescription.

We mention a few properties of (\ref{pwm}).
For $\qf \neq 0$, the limit
$\kf \rightarrow 0$ is singular \cite{MT}.
By explicit calculation one finds:
\beqn
E^{(\nu)} (\kf, \qf -\kf) =      2 \pi
     \delta^{(2)} (\kf) \left( \frac{1}{q^2} \right)^{\frac{1}{2}+i \nu}
\frac{\Gamma(1+2 i \nu)}{\Gamma(-2 i \nu)}
\frac{\Gamma(-\frac{1}{2}-i\nu) \Gamma(\frac{1}{2}-i\nu) }
  {\Gamma(\frac{3}{2}+i\nu)\Gamma (\frac{1}{2} + i \nu)}
           +O (\frac{\qf \cdot \kf}{\kf ^2} )
\label{dfu}
\eeqn
The delta-function term is dictated by the conformal invariance of the
BFKL-kernel, and it does not contribute if the BFKL-Pomeron is coupled
to an
external color singlet state which vanishes as $\kf \rightarrow 0$ (or
$\qf - \kf \rightarrow 0$). A useful regularization of the $\kf =0$ -
limit is obtained if in (\ref{pwm}) we introduce a nonzero
conformal dimension $\lambda$ for the (reggeized)
gluon field \cite{pol} :
\beqn
E^{(\nu,\lambda)}(\kf,\qf-\kf) &=&
                       \frac{4^{-i\nu}}{2\pi}\;
\frac{\Gamma(1+2i\nu)}{\Gamma(-2i\nu)}
\frac{\Gamma(-\frac{1}{2}-i\nu)}{\Gamma(\frac{3}{2}+i\nu)}
\frac{\Gamma^2(\frac{1}{2}-i\nu)}{\Gamma^2(\frac{1}{2}+i\nu)}\;
\nonumber \\
  & & \cdot  \int d^2 \rho_{1} d^2 \rho_{2}
        e^{i \kf \frho_1 + i (\qf - \kf)\frho_2}
      \frac{ (\rho_{12}^2)^{\frac{1}{2} -i \nu-\lambda}}
           { (\rho_1^2) ^{\frac{1}{2} -i \nu}
             (\rho_2^2) ^{\frac{1}{2} -i \nu} }
         \nonumber \\
     & = & 4^{1-\lambda}\pi\;
           \frac{\Gamma(1+2 i \nu)}{\Gamma(-2 i \nu)}
           \frac{\Gamma(-\frac{1}{2}-i \nu)}{\Gamma(\frac{3}{2}
                 +i \nu)}
           \frac{\Gamma(\frac{3}{2}+ i \nu - \lambda)
                 \Gamma(\frac{3}{2}- i \nu - \lambda)}
                        {\Gamma^2(\frac{1}{2}+i \nu)}
           \nonumber \\  &  &\cdot
           \int_0^1 d x x^{-\frac{1}{2} +i \nu}
                    (1-x)^{-\frac{1}{2} +i \nu}
[\qf^2 x(1-x)+(\kf-x\qf)^2]^{-\frac{3}{2} +i \nu +\lambda}
                   \nonumber \\
  &  &\cdot    _2F_1(\frac{3}{2}+i\nu-\lambda,
                  -\frac{1}{2}+i \nu +\lambda
                  ,1;\frac{(\kf-x\qf)^2}{\qf^2 x(1-x)+(\kf-x\qf)^2})
\label{gen}
\eeqn
For small $\kf^2$, one obtains :
\beqn
E^{(\nu,\lambda)}(\kf,\qf-\kf)
     & = & 4^{1-\lambda}\pi\;
           \frac{\Gamma(1+2 i \nu)}{\Gamma(-2 i \nu)}
           \frac{\Gamma(-\frac{1}{2}-i \nu)}{\Gamma(\frac{3}{2}
                 +i \nu)\Gamma^2(\frac{1}{2}+i \nu) }
                               \Gamma(1+2 i \nu)\Gamma(1-2 i \nu)
           \nonumber \\  &\cdot&   \! \! \! \! \! \! \!
          \left(\frac{1}{\qf^2}\right)^{\frac{1}{2}+i \nu}
 \! \! \!  (\kf^2)^{-1+\lambda} \! \!
          \left[\frac{i}{2 \nu}\frac{ \Gamma^2(1-\lambda)}
       {\Gamma(\frac{3}{2}+ i \nu-\lambda)\Gamma(-\frac{1}{2}
                                           -i \nu + \lambda)} +
           \mbox{c.c.} \right]
           +O (\frac{\qf \! \! \cdot \! \! \kf}{\kf ^2} )
\eeqn
which in the limit $\lambda \rightarrow 0$ leads us to
(\ref{dfu}), provided we identify :
\beqn
\lim_{\lambda \rightarrow 0}
2 ^{1-2\lambda}\left(\frac{1}{\kf^2}\right)^{1-\lambda} \cdot
\lambda =  \delta^{(2)}(\kf)
\eeqn
Finally, we note from (\ref{pom}) that, at fixed $\nu$, the
BFKL-pomeron factorizes in momentum space. This property is not present
in configuration space where the pomeron is a function of two
anharmonic ratios which link together the primed and unprimed
coordinates \cite{FGGP}. As a by-product of
this calculation, we find the conformal partial waves in a
mixed-representation which we give here for later use :
\beqn
E^{(\nu)}(\rho_{12},\qf) &=& \frac{1}{2}\;
\frac{\Gamma(1+2i\nu)}{\Gamma(-2i\nu)}
\frac{\Gamma(-\frac{1}{2}-i\nu)}{\Gamma(\frac{3}{2}+i\nu)}
\frac{1}{\Gamma^2(\frac{1}{2}+i\nu)}\;
(\qf^2)^{-i\nu} \,\rho_{12}
\nonumber \\ & &
\int_0^1 dx [x(1-x)]^{-\frac{1}{2}}\,e^{-i \qf \cdot \rho_{12} (1-x)}
K_{-2 i \nu}(|\qf||\rho_{12}|\sqrt{x(1-x)})
\label{mix}
\eeqn

Finally, for the (nonperturbative) coupling of the BFKL ladders to the
proton we use, as a guide for the dependence upon a hadronic scale
$Q_0^2$, the following simple model (at the point $ \qf =0$):
\beqn
V(k^2)=C \frac{k^2}{k^2 + Q_0^2}
\label{had}
\eeqn
where $Q_0^2$ denotes a hadronic scale of the order of $1 GeV^2$. As to
dependence upon $t$ and $\nu$, we shall assume that $C$ is a slowly
varying function.

Returning to (\ref{dcs}) and
putting together all these ingredients, we arrive at the following
expression for the partial wave $F$:
\beqn
F(\omega_1, \omega_2, t)&=&
  \int \frac{d^2 \lf}{(2 \pi)^3} \int \frac{d^2 \mf}{(2 \pi)^3}
  D_{(4;0)}^{(1;++)} (\lf,\qf-\lf,\mf,-\mf-\qf) \nonumber \\
& &  \cdot \int \frac{d^2 \lf'}{(2 \pi )^3}
           \Phi_{\omega_1} (\lf, \qf-\lf,\lf', \qf-\lf')
          V(\lf',\qf-\lf')  \nonumber \\
& &  \cdot \int \frac{d^2 \mf'}{(2 \pi )^3}
     \Phi_{\omega_2} (\mf,-\qf-\mf,\mf',-\qf-\mf') V(\mf',-\qf-\mf')
\label{pwf}
\eeqn
where $-\qf^2=t$ denotes the square of the momentum transfer, and
$\Phi$ and $V$ are given in (\ref{pom}), (\ref{had}), resp..

We now turn to (\ref{pwf}) and study its dependence upon $t$,
near $t=0$. We begin with the integrations in $\lf$ and $\mf$. For
$D_{(4;0)}$ we use (\ref{d01}), (\ref{d02}),
and from the two lower Pomerons we
have the two $E^{(\nu)}$-functions. This defines the vertex
$g_{\mu \mu_1 \mu_2}(q^2)$:
\beqn
&&\int \frac{d^2 \lf d^2 \mf}{(2 \pi)^6} \; \left[
(\lf^2)^{-\mu}  +((\lf-\qf)^2)^{-\mu}
+(\mf^2)^{-\mu} +((\mf+\qf)^2)^{-\mu}
-(\qf^2)^{-\mu}
-((\lf+\mf)^2)^{-\mu} \right. \nonumber\\
&& \left. -((\lf-\mf-\qf)^2)^{-\mu} \right]
\;E^{(\nu_1)} (\lf, \lf-\qf)  E^{(\nu_2)} (\mf, -\mf-\qf)
 \;=\;\frac{ (q^2)^{\mu_1 + \mu_2 - \mu} }{\mu_1 + \mu_2 -\mu} \;
g_{\mu \mu_1 \mu_2}(q^2)
\label{ver}
\eeqn
where we have used $\mu_i= -\frac{1}{2} - i\nu_i$.
In course of performing the integrals over $\lf$ and $\mf$ one finds
that, out of the seven terms on the lhs, the result will come from the
last two terms only; the other serve
as regulators in either the infrared or the ultraviolet region.
Rather than presenting details of the calculations, we shall
limit ourselves to a brief description of the major steps. First, for
the $E^{(\nu)}$ functions we use the representation (\ref{pwm}):
the hypergeometric
functions are written as a power series in their arguments, and we
consider term by term. It is convenient to perform the shifts
$\lf \rightarrow \lf-\qf z_1$, $\mf \rightarrow \mf+\qf z_2$
where $z_1$ and $z_2$ denote
the $x$-parameters of the representation (\ref{pwm}) for the two
BFKL-Pomerons. Now it is not difficult to perform the integration
over $\lf$: one ends up with a string of terms consisting of one
dimensional finite integrals involving hypergeometric functions
and powers. For the remaining integral over $\mf$ one observes that
convergence in the ultraviolet region holds as long as the condition
\beqn
\mu > \mu_1 + \mu_2
\eeqn
is satisfied ($\mu_i = -\frac{1}{2} - i \nu_i$), i.e. the $\mu$
integration contour is to the right of $\mu_1 + \mu_2$.
In the infrared region, $q^2$ serves as a regulator. Consequently, the
lhs of expression (\ref{ver}),
in the neighbourhood of $\mu=\mu_1 + \mu_2$,
behaves as
\beqn
\frac{ (q^2)^{\mu_1 + \mu_2 - \mu} }{\mu_1 + \mu_2 -\mu}
\cdot g_{\mu \mu_1 \mu_2}(q^2),
\label{pol}
\eeqn
and the remaining vertex $g_{\nu \nu_1 \nu_2}(q^2)$ has a finite limit
as $q^2 \rightarrow 0$:
\beqn
g_{\mu \mu_1 \mu_2}(0) \;=\;\frac{2 \pi ^2}{(2 \pi)^6}
    \frac{ \Gamma(1-\mu)}{\Gamma(\mu)}
  \frac{ \Gamma(\mu_1) \Gamma(\mu_2)}{ \Gamma(1-\mu_1) \Gamma(1-\mu_2)}
\label{lim}
\eeqn
We mention that (\ref{ver}) can also be evaluated at $q^2=0$
directly. The
result is
\beqn
(2 \pi) \delta(\mu_1 + \mu_2 - \mu) g_{\mu \mu_1 \mu_2}(0)
\label{del}
\eeqn
Both the pole in (\ref{pol}) and the $\delta$-function in (\ref{del})
express the "conservation of conformal dimension $\mu$" \footnote{
The proof of conformal invariance is under consideration.}
: in (\ref{del}) one
sees it directly, whereas for (\ref{pol}) we will show further below
that in the limit $q^2 \rightarrow 0$ the pole at $\mu= \mu_1 +
\mu_2$ will dominate. This result will be shown to have interesting
consequences for
the energy dependence of the inclusive cross section in the triple
Regge region at $t=0$.
Finally, we mention that (\ref{ver}) also contains terms without the
singularity shown in (\ref{pol}): one can show that for small $q^2$
they vanish faster that the contribution coming from (\ref{pol}).
Therefore,
they will not be considered here.

We return to the partial wave in (\ref{pwf})
and look at the implications of
our result for (\ref{ver}).
We still need to perform the integration over the
variables $\mu, \mu_1, \mu_2$, using the saddle point approximation.
The relevant terms are:
\beqn
\int \int \int \frac{d \mu}{2 \pi i} \;\frac{d \mu_1}{2 \pi i}
     \frac{d \mu_2}{2 \pi i} \;
\frac{(\frac{q^2}{Q_0^2})^{\mu_1 + \mu_2 - \mu}}{\mu - \mu_1 - \mu_2}
\frac{e^{y(\chi(\mu_1) + \chi(\mu_2))}}{\mu+1}
\left( \frac{Q^2}{Q_0^2} \right) ^{\mu}
\label{res}
\eeqn
where $y=\ln{1/x_B}$ and $\chi(\mu_i) = \chi(0, \nu_i)$.
The single pole at $\mu=-1$ arises from combining the double pole
in (\ref{d20}) with the zero in (\ref{lim}). As $q^2$ is small
we close the $\mu$-contour to the left, obtaining the two contributions
from the poles at $\mu=\mu_1 + \mu_2$ and $\mu=-1$:
\beqn
\int \int \frac{d \mu_1 d \mu_2}{(2 \pi i)^2}  \,\,
[
       \frac{\left( \frac{Q^2}{Q_0^2} \right) ^{\mu_1 + \mu_2}}
                           {\mu_1 + \mu_2 +1}
      - \frac{\left( \frac{q^2}{Q_0^2} \right) ^{1+ \mu_1 + \mu_2}
              \left( \frac{Q^2}{Q_0^2} \right) ^{-1}  }
                            {\mu_1 + \mu_2 +1}
             ] \,\,e^{y[\chi(\mu_1) + \chi(\mu_2)]}.
\label{twc}
\eeqn
We restrict ourselves to the case $\ln(Q^2/Q_0^2) \ll y$.
We begin with $q^2$ near $Q_0^2$ and perform the usual saddle
point analysis. The main
contribution comes from $\mu_1 = \mu_2 = - \frac{1}{2}$  (i.e.
$\mu_1 + \mu_2 +1$ is small), and the pre-exponent in (\ref{twc})
behaves as $\ln(Q^2/q^2)$:
\beqn
(\ref{twc}) \sim  \frac{Q_0^2}{Q^2} \, \ln{(\frac{Q^2}{q^2})}
           \left( \frac{1}{x_B} \right ) ^{2 \omega_{BFKL}}
       \frac{1}{2 \pi \ln(1/x_B) \chi''(-1/2)}
\label{smq}
\eeqn
with $\omega_{BFKL} = \chi(-1/2) = \frac{N_c \alpha_s}{\pi} 4
\ln 2$.

The $q^2$-dependence in (\ref{smq})
seems to indicate that the expression
diverges at
$q^2=0$, but this is not the case. Namely, when $\ln{Q_0^2/q^2}$ becomes
large (of the order of $y$ or even larger), the saddle point analysis
of (\ref{twc}) has to be modified.
Starting with the first of the two terms,
we introduce the variables $\mu_{+}=\mu_1 + \mu_2$ and
$\mu_{-} = \mu_1 - \mu_2$. For $\mu_{-}$ we again use the saddle point
approximation, whereas for $\mu_{+}$ we move the contour to the left
(parallel to the imaginary axis with real part $-1$, with a small
semicircle to the right of the point $\mu_{+} = -1$). The result of
the $\mu_{+}$-integral comes only from the semicircle and equals half
the residue (we note that the same result would have been obtained,
if we would have used (\ref{del}),
i.e. putting directly $t=0$ in (\ref{ver})).
As to the second term in (\ref{twc}),
the saddle point analysis now has
to take into account that there are two large parameters:
writing the $q^2$-factor as an exponential, we have $y$ and
$\ln{Q_0^2/q^2}$. With
growing $\ln{Q_0^2/q^2}$, the saddle point conditions become
\be
\chi'(\mu_1) = \chi'(\mu_2) = \frac{\ln{Q_0^2/q^2}}{y}
\ee
i.e. the saddle points of the $\mu_1, \mu_2$ integrals start
to move away from $-1/2$ more and more towards $\mu_1 = \mu_2 = 0$.
Consequently, in the limit $q^2 = 0$, the power of $q^2$ gets close to
unity, and the term vanishes.
Therefore, the small-$t$ limit of (\ref{twc})
comes only from the first term and equals:
\beqn
(\ref{twc})
=\frac{1}{4} \frac{Q_0^2}{Q^2} (\frac{1}{x_B})^{2 \omega_{BFKL}}
        \frac{1}{\sqrt{\pi \ln(1/x_B)
                             \chi''(-1/2)}}
             [ 1+ O(\sqrt{\frac{\ln ^2 (Q^2/Q_0^2}{\ln(1/x_B)
                                                     \chi''}}) ]
\label{smt}
\eeqn
This has to be compared with (\ref{smq}) which is valid only for
nonzero momentum transfer ($\ln(Q_0^2/q^2) \ll y$).
As a function of $q^2$, we expect to see a strong variation:
going from (2.25) to (2.27), the cross section grows but reaches
a finite limit. A similar saddle point analysis of the $q^2$-derivative
of (2.23) shows that the derivative tends to infinity as
$q^2 \rightarrow 0$:
the cross section therefore develops a cusp at $t=0$ (a more detailed
discussion will be given further below).

To complete our analysis of (\ref{pwf})
we have to couple the BFKL Pomerons
to the proton. The $\lf'$ integral in the second line of (\ref{pwf})
couples the second $E^{(\nu)}$-factor of the BFKL Pomeron to the
proton and, hence, belongs to nonperturbative physics. We assume that
the limit $q^2 \rightarrow 0$ is finite and has no strong variation
in the conformal dimensions $\mu_i$. Furthermore, guided by the simple
model (\ref{had}) we expect that the coupling should scale as
$(Q_0^2)^{-1/2 - i \nu_i}$. As a result, the coupling of the BFKL
Pomeron to the proton in the region of small $t$ is taken as:
\beqn
(Q_0^2)^{-1/2 - i \nu_i} \frac{C}{2 \sqrt{2}}
\eeqn
where $C$ is the constant from (\ref{had}),
independent of $\mu_i$ and $t$.

Before we present our final formula for the cross section, we
comment on the typical momentum scale at the upper end of the
two BFKL ladders. Starting from the hadron vertex at the lower end
where the average momentum lies in the vicinity of the hadronic scale
$Q_0^2$, we move upwards, and the distribution in transverse momentm
evolves in accordance with the diffusion mechanism. At first sight one
might expect that at the upper end the large scale of the photon mass
$\sqrt{Q^2}$ forces the diffusion into the ultraviolet direction: this
expectation, however, is not correct. Namely, if we consider the
quark loop as the first cell of a GLAP evolution ladder which
provides the biggest contribution to the cross section
only if the difference between the momentum scales at the
upper and lower end is as large as possible, then it becomes plausible
that we have a competition between the GLAP dynamics from the quark
loop and the BFKL diffusion mechanism from below. As a result, the
scale at the upper end of the ladders is pushed into the infrared
region. A computer analysis confirms this picture ~\cite{BV}.

Collecting finally all our results, we arrive at the following
expression for the cross section (\ref{dcs}) at $t=0$:
\be
\left. \frac{d \sigma^{DD}}{dt} \right|_{t=0} = \sum_f
       \frac{2 e_f^2 \alpha_{em} \alpha_s^2 C^2}
            {9 Q^2 Q_0^2} \frac{1}{\sqrt{21 \alpha_s \zeta(3) y}}
            (\frac{1}{x_B})^{2 \omega_{BFKL}}.
\ee

For comparison, we quote the result for the total cross section,
calculated with the BFKL-ladders and the same coupling to the proton
(Fig.3):
\beqn
\sigma(\gamma^* + \mbox{proton}) =
\sum_f \frac{e_f^2 \alpha_{em} \alpha_s C 9 \sqrt{2} \pi^3}
                             {64 \sqrt{Q^2 Q_0^2}}
             (\frac{1}{x_B})^{\omega_{BFKL}}
\frac{\exp(\frac{-\pi (\ln{Q_0^2/Q^2}) ^2}{4 \alpha_s 42 \zeta(3) y})}
     {\sqrt{42 \zeta(3) \alpha_s y}}
\label{tot}
\eeqn
For a phenomenological analysis one might think of taking the ratio
$\frac{d \sigma^{DD}}{dt}$ over $\sigma$ and determining the
unknown constant $C$ from a fit to $F_2$:
\be
\frac{1}{\sigma}\,
\left. \frac{d \sigma^{DD}}{dt} \right|_{t=0} \;=\;
\frac{1}{\sum_f e_f^2}\;\frac{2^{15}}{3^6\,\pi^4}\;
\frac{\sqrt{21 \zeta(3) \alpha_s y}}{Q^2}\;F_2
\label{com}
\ee
In eq.(\ref{com})
we have neglected corrections of the type $\ln(Q^2/Q^2_0)$.
Taking $x_{Bj}=10^{-3}$ and $Q^2=10GeV^2$ we find, as a quantitative
prediction, that 10 \% of the usual DIS-events are
diffractively produced $q \bar{q}$-pairs. This is clearly
only a rough estimate, and its validity is restricted by the condition
that the BFKL Pomeron has to be applicable. This excludes configurations
where one of the quarks is soft, but, nevertheless, it turns out to be
a reasonable value.

\section{The Configuration Space Representation}
\setcounter{equation}{0}

In this section we present an alternative way of deriving the cross
section for the diffractive $q\bar{q}$-production.
We consider the $q\bar{q}$-final state as two opposite colour charges
in the configuration or impact parameter space. The important parameter
is their separation ${\bf r}$.
This representation as Colour Dipole in the impact parameter space
(see for example ref. \cite{Mue,NZ}) has the nice property that
it diagonalizes the scattering matrix in the limit
of high energy (small $x$) and small momentum transfer, i.e the
impact parameter is a good quantum number. Multiple
scattering turns out to be simply the product of single scattering due
to which the calculation can be performed in a compact way
and shows from the beginning the final factorized
form as in eq. (\ref{ver}).
We normalize the wave function $\Psi({\bf r})$
using $D_{(2;0)}$ which was introduced in ref. \cite{BW}:
\beqn
D_{(2;0)}({\bf k})&=&\int d^2{\bf r}\;|\Psi({\bf r})|^2\;
                        (1-e^{i{\bf k} \cdot {\bf r}})\;
   \nonumber            (1-e^{-i{\bf k} \cdot {\bf r}})  \\
          &=&\int d^2{\bf r}\;|\Psi({\bf r})|^2\;
             (2-e^{i{\bf k}\cdot{\bf r}}-e^{-i{\bf k}\cdot{\bf r}})
\label{wvf}
\eeqn
with
\be
|\Psi({\bf r})|^2\;=\;
\sum_f   e_f^2\alpha_s\,\frac{\sqrt{8}}{4\,\pi^2}\;
                                  \int_0^1 d\alpha\;
\left[1-2\alpha(1-\alpha)\right]\;\alpha(1-\alpha)Q^2
\;K_1^2\left(\sqrt{\alpha(1-\alpha)Q^2r^2}\right)\;\;.
\ee
\noindent
$K_1$ is the modified Bessel function of first order. For comparision
see ref. \cite{Mue} and \cite{NZ}.
Next, we would like to generalize from two gluons to four gluons with
each of the gluon pairs (1,2) and (3,4) in the colour singlet state.
We can apply the same wave function as in eq.(\ref{wvf}).
One only needs
to add two more factors of the type $(1-e^{i{\bf k}\cdot {\bf r}})$,
the corresponding colour factor and a $g^2$ for the coupling
of two more gluons. Accordingly, we can rewrite $D_{(4;0)}$ as:
\beqn
D_{(4;0)}({\bf k})\;=\;g^2\frac{\sqrt{2}}{3}\;
                       \int d^2{\bf r}\;|\Psi({\bf r})|^2
                        (1-e^{i{\bf k_1} \cdot {\bf r}})\;
                        (1-e^{i{\bf k_2} \cdot {\bf r}}) \;
                        (1-e^{i{\bf k_3} \cdot {\bf r}})\;
                        (1-e^{i{\bf k_4} \cdot {\bf r}})
\label{wvf2}
\eeqn
with ${\bf k_1}+{\bf k_2}+{\bf k_3}+{\bf k_4}=0$. As before, we fix
the momentum tranfer along the Pomeron ${\bf q}={\bf k_1}+{\bf k_2}=
-{\bf k_3}-{\bf k_4} $. The notation for the internal transverse momenta
of the left and right Pomeron were chosen to be ${\bf l}={\bf k_1}$ and
${\bf m}={\bf k_3}$. With this notation eq.(\ref{wvf2}) gives:
\beqn
D_{(4;0)}({\bf k})\;=\;g^2\frac{\sqrt{2}}{3}
                     \;\int d^2{\bf r}\;|\Psi({\bf r})|^2\;
&&\left[1-e^{i{\bf l} \cdot {\bf r}}\,
 +\,(1-e^{-i{\bf l} \cdot {\bf r}})e^{i{\bf q}\cdot{\bf r}}\right]
 \;\cdot \label{wvf3} \\
\;\cdot&&\left[1-e^{i{\bf m} \cdot {\bf r}}\,   \nonumber
+\,(1-e^{-i{\bf m} \cdot {\bf r}})e^{-i{\bf q}\cdot{\bf r}}\right]\;\;.
\eeqn
This expression is already factorized corresponding to each
of the Pomerons.

It was shown in section 2 that with the help of the
Pomeron-eigenfunction $E^{(\nu)}$ for a given momentum transfer
${\bf q}$ the solution of the Lipatov equation can be factorized.
The vertex $g$ (see eq.(\ref{ver}))
is part of the projection of $D_{(4;0)}$
on these eigenfunctions.
In the following we will use the mixed representation (\ref{mix}) of the
Pomeron-eigenfunction $E^{(\nu)}$ and take the Fourier transformation
of $D_{(4;0)}$.
It is enough to look at one of the factors of (\ref{wvf3})
e.g. $[1-e^{i{\bf l} \cdot {\bf r}}\,
     +\,(1-e^{-i{\bf l} \cdot {\bf r}})e^{i{\bf q}\cdot{\bf r}}]$.
Its Fourier transformed is
$\delta({\bf \rho})-\delta({\bf r}+{\bf \rho})\,+
\,[\delta({\bf \rho})-\delta({\bf r}-{\bf \rho})]\,
e^{i{\bf q}\cdot{\bf r}} $ where the impact parameter
${\bf \rho}$ corresponds to ${\bf l}$. A crucial
property of this expression is its vanishing after the integration
over ${\bf \rho}$. This reflects the colour cancellation and is
an important requirement to restore the conformal invariance as was
shown in \cite{Lip}. Inserting the eigenfunction
$E^{(\nu)}({\bf \rho},{\bf q})$
and integrating over ${\bf \rho}$ we end up with
\beqn && g^2\frac{\sqrt{2}}{3\,\pi^2}
\;\int d^2{\bf r}\;|\Psi({\bf r})|^2\;\;
E^{(\nu_1)}({\bf r},{\bf q})\;E^{(\nu_2)}({\bf r},-{\bf q}) \\
&=&g^2\frac{\sqrt{2}}{3\,\pi^2}
    \;\int dr\; r\;|\Psi(r)|^2\;\int_0^{2\pi} d\phi\;
E^{(\nu_1)}({\bf r},{\bf q})\;E^{(\nu_2)}({\bf r},-{\bf q})\;\;.
                               \nonumber
\eeqn
Note that $E^{(\nu)}$ vanishes when ${\bf r}$ equals zero, and we have
made use of the relation $E^{(\nu)}({\bf r},{\bf q})\,e^{i{\bf q}\cdot
{\bf r}}$ $=\;E^{(\nu)}(-{\bf r},{\bf q})$. In order to recover the
vertex $g$ in terms of the three conformal dimensions $\nu,\nu_1$ and
$\nu_2$ we have to take the Mellin transformation of the wave function:
\beqn
&&\int dr \, r\;|\Psi(r)|^2\;r^{1+2i\nu} \nonumber \\&=&\nonumber
\sum_f   e_f^2\alpha_s\,\frac{\sqrt{8}\,4^{i\nu}}{16\pi}\;
\frac{\Gamma(5/2+i\nu)}{\Gamma(2+i\nu)}\frac{\Gamma(1/2+i\nu)}{1/2+i\nu}
\frac{\Gamma(1/2-i\nu)}{1/2-i\nu}\frac{\Gamma(5/2-i\nu)}{\Gamma(2-i\nu)}
\frac{\Gamma(3/2+i\nu)}{-\Gamma(-1/2-i\nu)} \\
&=&\frac{4^{i\nu}}{2\pi}\;\frac{\Gamma(3/2+i\nu)}
{-\Gamma(-1/2-i\nu)}\;\;\tilde{D}_{(2;0)}(-1/2-i\nu)
\eeqn
$\tilde{D}_{(2;0)}$ is the Mellin transformed of $D_{(2;0)}$ in the
momentum space (eq.(\ref{d20})).
The factor in front of $\tilde{D}_{(2;0)}$
is the inverse of a factor which follows from
the Fourier transformation of $(k^2)^{-3/2-i\nu}$.
This relation is illustrated by taking $E^{(\nu)}$
in the forward direction, i.e. at ${\bf q}=0$ (see the discussion after (2.9)):
\be
E^{(\nu)}({\bf \rho},{\bf q}=0)
\;=\;\sqrt{2}\;4^{-i\nu-1}\;
\frac{-\Gamma(-1/2-i\nu)}{\Gamma(3/2+i\nu)}\;\rho^{1+2i\nu}
\ee
The vertex $g$ which was defined in eq.(\ref{ver}) can now be rewritten
in terms of the impact parameter ${\bf r}$:
\beqn
&&\nonumber g_{\nu\nu_1\nu_2}(q^2)\;
\frac{q^{-1+2i\nu-2i\nu_1-2i\nu_2}}{1/2-i\nu+i\nu_1+i\nu_2} \\
&=&\frac{4^{i\nu}}{2\pi^3}\;\frac{\Gamma(3/2+i\nu)}
{-\Gamma(-1/2-i\nu)}\;\int dr \;r^{-2i\nu-2}\,\int_0^{2\pi} d\phi\;
E^{(\nu_1)}({\bf r},{\bf q})\;E^{(\nu_2)}({\bf r},-{\bf q})\;\;.
\label{ver2}
\eeqn

We are mainly interested in the limit $q^2\rightarrow 0$ of expression
(\ref{ver2}).
Following the discussion of section 2 we have to evaluate the
residue at the point $1-2i\nu + 2i\nu_1 + 2i\nu_2=0$, e.g.:
\be
\left. g_{\nu\nu_1\nu_2}(q^2)
\frac{q^{-1+2i\nu-2i\nu_1-2i\nu_2}}{1/2-i\nu+i\nu_1+i\nu_2}
\;\right|_{q^2=0} \;=\;
2\pi\,\delta(1/2-i\nu+i\nu_1+i\nu_2)\;g_{\nu\nu_1\nu_2}(0) \nonumber\\
\ee
Finally we can evaluate the vertex $g_{\nu\nu_1\nu_2}$ by partial
integration:
\beqn
g_{\nu\nu_1\nu_2}(0)  &=&
\;\frac{4^{i\nu}}{2\pi^3}\,\frac{\Gamma(3/2+i\nu)}
{-\Gamma(-1/2-i\nu)}
\int dr \;(1-2i\nu+2i\nu_1+2i\nu_2)\,r^{-2i\nu+2i\nu_1+2i\nu_2}
\;\;\cdot\nonumber\\ &&\cdot\;\;
\int_0^{2\pi} d\phi \;\;r^{-1-2i\nu_1}E^{(\nu_1)}({\bf r},{\bf q})\;
\;r^{-1-2i\nu_2}E^{(\nu_2)}({\bf r},-{\bf q})\\\nonumber
&=&
\;\frac{1}{\pi^2}\;4^{i\nu}\,\frac{\Gamma(3/2+i\nu)}
{-\Gamma(-1/2-i\nu)}\;\;
\left[r^{-1-2i\nu_1}E^{(\nu_1)}({\bf r},{\bf q})\right]_{{\bf r}=0}\;
\;\left[r^{-1-2i\nu_2}E^{(\nu_2)}({\bf r},-{\bf q})\right]_{{\bf r}=0}
\\\nonumber &=& \frac{1}{8\,\pi^2}\;
\frac{\Gamma(3/2+i\nu)}{-\Gamma(-1/2-i\nu)}
\frac{\Gamma(-1/2-i\nu_1)}{\Gamma(3/2+i\nu_1)}
\frac{\Gamma(-1/2-i\nu_2)}{\Gamma(3/2+i\nu_2)}
\eeqn
The final result agrees with (2.21).

The main advantage of this derivation is its compact form and
the absence of subtraction terms which make the calculation in the
momentum space more involved. But, the calculation can only be
performed at finite ${\bf q}$ which serves as infrared cutoff whereas
the momentum integrals in section 2 are finite at ${\bf q}=0$.

\section{The Triple Pomeron Vertex}
\setcounter{equation}{0}

We return to momentum space and consider the more general case of the
inclusive cross section in the triple Regge region (Fig.4):
\beqn
\frac{d^2 \sigma}{dt dM^2} = \frac{1}{16 \pi M^2}
                             \int \frac{d \omega }{2 \pi i}
                             \int \frac{d \omega_1 }{2 \pi i}
                             \int \frac{d \omega_2 }{2 \pi i}
      \left(\frac{s}{M^2}\right)^{\omega_1 + \omega_2}
      \left(\frac{M^2}{Q^2}\right)^{\omega}
      F(\omega, \omega_1, \omega_2, 0,t,t)
\eeqn
where $M$ denotes the invariant mass of the diffractively produced
system.
The analytical calculation of the partial wave $F$ has been done in
{}~\cite{BW}, and in this paper we study the change in the energy
dependence near $t=0$. As the dependence upon $s/M^2$ and $M^2/Q^2$
is determined by the $\omega$-singularities in the two lower legs
and the upper t-channel, resp., we expect $\omega_1 =
\omega_2 = \omega_{BFKL}$, and $\omega = \omega_4$ or $\omega =
\omega_{BFKL}$  ($\omega_4$ denotes the leading singularity of the
four-gluon state).
In ~\cite{BW} it was shown that the full inclusive cross section
comes as a sum of two terms: in the first term (Fig.1b), the two
lower BFKL ladders couple, via a disconnected vertex, to the upper
BFKL ladder; consequently we expect $\omega = \omega_{BFKL}$.
In the second term (Fig.1a), the two lower BFKL ladders
first merge into a four-gluon state, where the four gluon lines
interact in all possible ways; then there is a transition vertex
from the four-gluon state to the two-gluon state, which has the
familiar BFKL interaction kernels and connects to the fermion box at the
top of the diagrams. As a result we have contributions from both
$\omega=\omega_{BFKL}$ and $\omega=\omega_4$.

The following discussion we
will show that these expectations for $\omega$ (and therefore the
$M^2$-dependence) are not correct when
$t \rightarrow 0$: as a result of the conservation of conformal
dimension found in (\ref{pol}) and (\ref{del}),
there is no coupling between the
leading $\omega$-singularities in all three channels, i.e. the
coupling between the three BFKL-singularities generated by the
three ladders in Fig.1b vanishes at $t=0$.
In Fig.1a our ignorance of the $\omega$-singularity of the four-
gluon state prevents us from carrying out a complete analysis:
presently we can only conclude that the conservation of the conformal
dimension holds and that the leading singularity (whatever it will be)
decouples from the lower two BFKL-singularities. In any case, we
predict a change in the $M^2$-dependence near $t=0$.

In analogy with the structure of the inclusive cross section, our
following analysis goes in two steps. First we consider the first part,
i.e. the direct coupling of three BFKL ladders.
This part will be referred to as the ``triple ladder vertex``,
in order to distinguish this part from the full ``triple Pomeron
vertex``.
In the second part we turn to the more complex case where the
the two BFKL-ladders couple to the four-gluon state; here our analysis
will remain somewhat incomplete.

\subsection{The Triple Ladder Vertex}

The expression for Fig.1b can be obtained from (\ref{pwf}) by simply
replacing the fermion loop $D_{(4;0)}$ by the full sum of gluon ladders
$D_{(4)}^R$ (in the notation on ~\cite{BW}).
In analogy with (\ref{d01})
$D_{(4)}^R$ can also be written as a sum of seven terms: on the lhs
of (\ref{d01}), we replace $D_{(4;0)}$ by $D_{(4)}^R$, and on the rhs
all the $D_{(2;0)}$'s by the corresponding $D_{(2)}$'s. As reviewed in
{}~\cite{B}, the two-gluon function $D_{(2)}(\kf, -\kf)$ has a
representation analogous to (\ref{d01}). Singularities
in the $\mu$-plane
are poles which lie at distances of order $g^2$ away from positive and
negative integer $\mu$-values. The integrations over $\lf$ and $\mf$
are done in the same way as described before ((\ref{pol})),
leading again
to (\ref{pol}) and (\ref{del}) for $t \neq 0$ and $t=0$, resp.
The evaluation of
the remaining $\mu$-integrals, however, is slightly different. Namely,
instead of (\ref{res}), we need to calculate
\beqn
\int \int \int \frac{d \mu}{2 \pi i} \frac{d \mu_1}{2 \pi i}
     \frac{d \mu_2}{2 \pi i}
\frac{\left( \frac{q^2}{Q_0^2} \right)^{\mu_1 + \mu_2 - \mu}}
          {\mu - \mu_1 - \mu_2}  \left( \frac{Q^2}{Q_0^2} \right)^{\mu}
\, \, e^ {y_M \chi(\mu) + y_s [ \chi(\mu_1) + \chi(\mu_2)]}
\label{4res}
\eeqn
where $y_M = \ln{M^2/ Q^2}$ and $y_s = \ln{ s/M^2}$. We consider
the low mass region $y_M \ll y_s$ and take $\ln(Q^2/Q_0^2)\ll y_s$.
The most interesting case is $y_M \leq \ln(Q^2/Q_0^2)$.
We begin with $q^2$ being of the order $Q^2$, where we simply
repeat the standard saddle point analysis:
one finds $\mu_{1S} = \mu_{2S} = -1/2$ and, as a condition on $\mu_S$,
\beqn
0= y_M \chi'(\mu_S) + \ln{Q^2/q^2}.
\eeqn
Since for $q^2$ near $Q^2$ $\ln{Q^2/q^2}$ will be small compared to
$y_M$, we obtain
\beqn
\mu_{S} = -\frac{1}{2} - \frac{ \ln{Q^2/q^2} }{ y_M \chi''(
-\frac{1}{2})}.
\eeqn
The result for (\ref{4res}) is:
\beqn
2 \frac{Q_0^2}{Q^2}\;
             \left( \frac{q^2}{Q^2} \right) ^{-1/2}
             \left( \frac{s}{M^2} \right) ^{2 \omega_{BFKL}}
             \left( \frac{M^2}{Q^2} \right) ^{\omega_{BFKL} }
\frac{\exp(-\frac{\ln^2(Q^2/q^2)}{2y_M \chi''})}
        {\sqrt{2 \pi \chi''(-\frac{1}{2}) y_M }}
\frac{\exp(-\frac{\ln^2(q^2/Q_0^2)}{y_s \chi''})}
       {2 \pi \chi''(-\frac{1}{2}) y_s }
\label{4smq}
\eeqn
Here the $q^2$-dependence agrees with what first has been found in
{}~\cite{MueP}. Moving now to smaller $q^2$-values, we come into the
region near $q^2 = Q_0^2$ where $y_M \leq \ln Q^2/q^2 \ll y_s$, and
the  solution to (4.3) is no longer given by (4.4). Instead,
$\mu_S$ moves close to $-1$:
\beqn
\mu_{S} = -1- \sqrt{\frac{\alpha_s N_c y_M}{\pi \ln{Q^2/q^2} }}
\eeqn
which leads to the following result for (\ref{4res}):
\beqn
 \frac{Q_0^2}{Q^2}\;
     \exp{\left( 2 y_s \omega_{BFKL} +
   2  \sqrt{\frac{N_c \alpha_s}{\pi} y_M \ln{\frac{Q^2}{q^2}}}\right)}
 \;\left(16 \pi N_c \alpha_s y_M
           \ln{\left(\frac{Q^2}{q^2}\right)}\right)^{-\frac{1}{4}}
\;\frac{\exp(-\frac{\ln^2(q^2/Q_0^2)}{y_s \chi''})}
          {2 \pi \chi''(-\frac{1}{2}) y_s }
\eeqn
We thus obtain, for the $M^2$-dependence, the typical GLAP result for
the upper ladder with $q^2$ as lower scale and starting point of the
evolution.

Before we come to the third region $0 < q^2 < Q_0^2$ we mention,
for completeness, also the other case, $\ln(Q^2/Q_0^2) \ll y_M$.
For $q^2$ near $Q^2$, the result of (4.2) is, again, given by (4.4);
when $q^2$ moves close to $Q_0^2$, the saddle point $\mu_S$ stays
near $-1/2$ (eq.(4.4)), and (4.5) remains valid.

Finally the limit $q^2 \rightarrow 0$ ($q^2 < Q_0^2$).
When in (\ref{4smq})
the momentum transfer $q^2$ becomes smaller and smaller,
we observe a similar phenomenon as described after (\ref{smq}):
the effective power of
$q^2$ increases from $-1/2$ to zero, and the limit $t=0$ is finite. To
see this in detail, we write $\left( \frac{q^2}{Q_0^2} \right)^{\mu_1 +
\mu_2 - \mu} = \exp{[ (\mu - \mu_1 - \mu_2) \ln{Q_0^2/q^2} ]}$, and we
search for the saddle point of the function:
\beqn
\psi(\mu, \mu_1, \mu_2) = (\mu - \mu_1 - \mu_2) \ln{Q_0^2/q^2}
            + y_M \chi(\mu) + y_s [ \chi(\mu_1) + \chi (\mu_2) ]
            + \mu \ln{Q^2/Q_0^2}
\eeqn
The conditions are:
\beqn
0 & = & \ln{Q^2/q^2} + y_M \chi'(\mu_S)  \\
0 & = & -\ln{Q_0^2/q^2} + y_s \chi'(\mu_{iS} ),
\eeqn
and one sees that for very small $q^2$ all saddle point values start
to move: $\mu_S$ moves to the left (towards the point $\mu= -1$
where the function $\chi$ becomes infinite with a negatice slope),
whereas the $\mu_{iS}$ start to move in the right direction $\mu_{iS} >
-1/2$ (at zero $\chi(\mu_i)$ tends to infinity with a positive slope).
(4.9) and (4.10) also indicate at which $q^2$-values the motion starts.
For the $\mu_i$, we need $\ln(Q^2/q^2)$ to become of the order $y_M$
(which in case of $y_M \leq \ln(Q^2/Q_0^2)$ has already been reached for
$Q_0^2 < q^2$, see above), whereas for $\mu$ the condition is
$\ln(Q_0^2/q^2) \approx y_s$. In order to evaluate the integral (4.2),
we shift the $\mu_i$-contours to the right (such that it always passes
through the saddle point), and we deform the $\mu$-contour as shown
in Fig.5. For $q^2 \rightarrow 0$ the saddle point approaches
$\mu=-1$, $\mu_i=0$ and the final result splits
(\ref{4res}) into two contributions: the pole contribution which
is independent of $q^2$, and a $q^2$-dependent part which can be
computed from the saddle points in eqs. (4.9) and (4.10).
Since we look for saddle points close to $\mu_i=0$ and $\mu=-1$,
we can approximate the $\chi$-function by its leading poles
and find:
\be
- \;\frac{q^2}{Q^2} \;\frac{\exp{\left(2 \sqrt{\frac{\alpha_s N_c}{\pi}
              y_M
           \ln{\frac{Q^2}{q^2}}}\right)}}{2 \left[\pi \alpha_s N_c
                y_M \ln{\frac{Q^2}{q^2}}\right]^\frac{1}{4}}\;
      \frac{\exp{\left(4 \sqrt{\frac{\alpha_s N_c}{\pi} y_s
           \ln{\frac{Q_0^2}{q^2}}}\right)}}{4 \sqrt{\pi \alpha_s N_c
                y_s \ln{\frac{Q_0^2}{q^2}}}}
\ee
(we remind that the starting point of our discussion, eq.(4.2),
represents a somewhat simplified approximation to the partial wave
in (4.1). In particular, the triple ladder vertex in (2.20)
is valid only near the point $\mu=\mu_1+ \mu_2$. The full expression
in (2.18) contains other poles at $\mu=\mu_1+\mu_2 -1,...$ which
lead to contributions that are nonleading at small $q^2$. Since
(4.11) represents such a nonleading contribution, we should, in
principle, have started from an approximation which is more
accurate than (4.2) and contains also the pole at
$\mu=\mu_1+\mu_2-1$. However, the modification of (4.11) consists only
of powers of $y_s, y_M$ and $\ln(Q_0^2/q^2)$ in front of the
exponential and is not essential for our discussion).
Eq. (4.11) shows the familar double leading log result:
$\ln(M^2/Q^2)\ln(Q^2/q^2)$ for the upper ladder ($\mu \simeq -1$)
with the usual ordering of the internal transverse momenta from
the large scale $Q^2$ at the top down to the small scale $q^2$,
whereas the lower ladders dependent on the double logs
$\ln(s/M^2)\ln(Q_0^2/q^2)$ with an inverse ordering ($\mu_i \simeq 0$)
from the lower scale $q^2$ at the vertex up to the 'larger' scale
$Q_0^2$ at the hadronic side of the diagram. It is easy to estimate
the typical scale of transverse momenta at the triple ladder vertex:
the contribution (4.11) will reach its maximum if
the $k_T^2$-evolution inside the ladders above and below are
as long as possible: consequently, the momenta at the triple
Pomeron vertex will try to be as small as possible, i.e. of the order
of $q^2$. This value should be rather independent of the energies.
Although (4.11), by itself, vanishes as $q^2 \rightarrow 0$, it is
nevertheless of physical interest since it determines the slope
of the cross section  near $t=0$. The derivative of (4.11) becomes
infinite at $t=0$; for finite (but small) $t$ it increases with both
$y_s$ and $y_M$. It should, however, be kept in mind that
this large slope near $t=0$, however, is due to very small momenta at
the triple ladder vertex. In this region, predictions which are
based upon a perturbative analysis are not reliable.

At $q^2 =0$ expression (4.11) vanishes and only the second
contribution, coming from the pole
in the $\mu$-plane at $\mu=\mu_1 + \mu_2$ gives a nonzero
contribution (the exponent of $q^2$ in (\ref{4res}) equals zero).
The integrals over $\mu_1$ and $\mu_2$ are done using the
saddle point approximation of the exponent:
\beqn
\psi(\mu_1, \mu_2) = y_M \chi(\mu_1 + \mu_2) + y_s [\chi(\mu_1)
                + \chi(\mu_2) ] + (\mu_1 + \mu_2) \ln{Q^2/Q_0^2}
\label{exp}
\eeqn

We consider a few cases which we can treat analytically. The most
interesting one is the low mass region $y_M \ll y_s$ which we have
discussed also for $q^2\neq 0$. The saddle point analysis
distinguishes between the two regions:
\beqn
(a) \,\,& y_s^2 y_M & \gg \left( \ln{ \frac{Q^2}{Q_0^2} } \right)^3 \\
(b) \,\,& y_s^2 y_M & \ll \left( \ln{ \frac{Q^2}{Q_0^2} } \right)^3
\eeqn
The saddle points are :
\beqn
(a) \,\,& \mu_{1S} = \mu_{2S} =& - \frac{1}{2}
+\left(\frac{N_c\alpha_s}{4 \pi}\frac{1}{\chi''(-\frac{1}{2})}
\frac{y_M}{y_s}\right)^{\frac{1}{3}} \\
(b) \,\,& \mu_{1S} = \mu_{2S} =& - \frac{1}{2}
+\sqrt{\frac{N_c\alpha_s}{4 \pi}\frac{y_M}{\ln \frac{Q^2}{Q_0^2}}},
\label{sp2}
\eeqn
and we obtain the following results for (\ref{4res}):
\beqn
&(a)&\,\, \frac{Q_0^2}{Q^2}\exp\left(2y_s\omega_{BFKL}+ \frac{3}{2}
\left(\frac{N_c\alpha_s y_M}{\pi}\right)^{\frac{2}{3}}
\left[\frac{ y_s \chi''(-\frac{1}{2})}{2}\right]^{\frac{1}{3}}\right)
\frac{1}{\sqrt{12}\pi y_s \chi''(-\frac{1}{2})} \label{sp1}\\
&(b)&\,\, \frac{Q_0^2}{Q^2}\exp\left(2y_s\omega_{BFKL} \,+\,2\,
\sqrt{\frac{N_c\alpha_s}{\pi}y_M \ln \frac{Q^2}{Q_0^2}} \right)
      \frac{1}{\sqrt{4\pi^2y_s\chi''(-\frac{1}{2})}}\left(\frac{y_M}
{\ln^3\frac{Q^2}{Q_0^2}}\frac{N_c\alpha_s}{16 \pi}\right)^{\frac{1}{4}}
\eeqn

Case (a) is the one which should be compared to our discussion for
for $q^2 \neq 0$ (\ref{4smq}) and $y_M\leq \ln(Q^2/Q_0^2)$. We compare
the results (4.5), (4.7), and (4.17), keeping the variables $y_s$,
$y_M$, and $Q^2/Q_0^2$ fixed. Going from the point $q^2=Q^2$ in (4.5)
to $q^2=Q_0^2$ in (4.7), we observe an increase (coming from the
exponential) which depends upon the ratio $\ln(Q^2/Q_0^2) / y_M$. At
$q^2=0$, (4.17) gives a finite value which, once more,
is larger than (4.7). This last increase will depend upon $y_s$:
the larger $y_s$, the stronger the enhancement. The change from
the behavior (4.7) to (4.17) takes place when $\ln(Q_0^2/q^2)$ is of
the order of $y_s$ (see above).

Let us discuss some characteristic features of this pole contribution
to (4.2). A typical property of the small mass region ($y_M \ll y_s$) is
the location of the saddle point of $\mu = \mu_1+\mu_2$ close to -1.
In this region the pole term of the first $\chi$-function
in eq. (\ref{exp})
dominates, i.e the transverse momenta of the upper ladder are strongly
ordered. So the physics inside the upper ladder is the same as discussed
after (4.11): the internal transverse momentum decreases (strongly
ordered) as we move down from the photon to the triple ladder vertex.
The two lower ladders, on the other hand, remain in the BFKL-region
($\mu_1$ and $\mu_2$ are close to $-1/2$), and we have the usual
diffusion in $\ln(k_T^2/Q_0^2)$ inside the ladders.
If we ask for the typical momentum scale at the triple ladder vertex,
we recognize two competing effects: the upper ladder with its strong
ordering tries to have as much evolution as possible, i.e. tends
to lower the momentum at the triple ladder vertex. From below, on the
other hand,
we have two (approximately) gaussian distributions in $\ln(Q_0^2/k_T^2)$
which are centered at the hadronic scale $Q_0^2$, have a width
of the order $y_s$ and, in particular, suppress the region of very
small momenta. Convoluting these gaussians with the
distribution from above, we find a maximum at $\ln(Q_0^2/k_T^2)=
(y_M y_s^2)^{\frac{1}{3}} +\frac{2}{3} \ln(Q^2/Q_0^2)$. Consequently,
for large energies
the average momentum value at the triple ladder vertex is small and
decreases with the energy:
\beqn
    <k_T^2/Q_0^2>  \sim exp(-const \cdot (y_My_s^2)^{\frac{1}{3}} ).
\eeqn
One therefore expects that with increasing energy the diffusion enters
more and more into the infrared region where perturbation theory
becomes unreliable. The two lower ``hard`` Pomerons, therefore, from
which our analysis had started, should more and more transform
themselves into ``soft`` (nonperturbative) ones.

Summarizing our discussion of case (a), our analysis of (4.2)
(which is proportional to the inclusive cross section) consists
of two pieces: the leading one (4.17) determines the size of the cross
section at $t=0$. The  first nonleading term (4.11) vanishes at $t=0$,
but it determines the slope which becomes infinite at $t=0$.
So our cross section, as a function of $t$, has a cusp at $t=0$.
Since the slope increases with $y_s$, the width of the cusp
shrinks with increasing energy.
This situation is reminiscent of the two gluon exchange discussed
in ~\cite{LR}: the result for the cross section is finite, but the
slope at $t=0$ is infinite, too. But our result differs from the simpler
case of ~\cite{LR} in that our cusp has an energy dependent width.
The physical origin for this kind of ``shrinkage`` lies in the
exponential factors in (4.11), i.e. the GLAP evolution above and
below the triple ladder vertex. The analysis of the typical momentum
scale at the triple ladder vertex indicates that, within our purely
perturbative analysis, more credibility should be given to
the value of the cross section at $t=0$ rather than to its slope near
$t=0$ which has been found to be a large distance effect.

The case (b) does not require much new discussion: at the triple
ladder vertex we again have the competition between the GLAP ordering
from above and the diffusion from below. The very large photon mass
now tries to pull the momenta towards large values; but it turns out
that the momentum scale at the central vertex never exceeds the
hadronic scale $Q_0^2$:
\be
<\ln(k_T^2) > \sim \ln(Q_0^2).
\ee
So the lower Pomerons again like to become nonperturbative, although
somewhat less than in the case (a).

Finally we also mention a few results on the large mass region near
$t=0$. We consider the two cases:
\beqn
  (c) &\,\,  y_M=y_s \gg \ln \frac{Q^2}{Q_0^2}   \\
  (d) &\,\,  y_M \gg y_s \gg \ln \frac{Q^2}{Q_0^2}
\eeqn
In the first case, we find the stationary point near $\mu_1= \mu_2
=-1/3$ and $\mu=-2/3$, in the second case near $\mu_1 = \mu_2 = -1/4$
and $\mu=-1/2$. The conditions are:
\beqn
(c)& \,\, \mu_{1S} = \mu_{2S} =& -\frac{1}{3}   \\
(d)& \,\,  \mu_{1S} = \mu_{2S} =& - \frac{1}{4}\;-\;
\frac{y_s \chi'(-\frac{1}{4})}{2y_M\chi''(-\frac{1}{2})}
\eeqn
They lead to the results:
\beqn
(c)& \,\, \left(\frac{Q_0^2}{Q^2}\right)^{\frac{2}{3}}\exp\left(
y_M\chi(-\frac{2}{3})+2y_s\chi(-\frac{1}{3})
\right)
\frac{1}{\sqrt{4\pi^2y_s\chi''(-\frac{1}{3})}}\;
\frac{1}{\sqrt{y_s\chi''(-\frac{1}{3})+2 y_M\chi''(-\frac{2}{3})}}
\\
(d)& \,\,
\sqrt{\frac{Q_0^2}{Q^2}}\exp\left(y_M\omega_{BFKL}+\,2y_s
\chi (-\frac{1}{4})\right)
\frac{1}{\sqrt{8\pi^2y_My_s\chi''(-\frac{1}{2})\chi''(-\frac{1}{4})}}.
\eeqn
The values of the $\chi$-functions are:
\beqn
(c) & \,\, \chi(-\frac{1}{3})\;=\;\chi(-\frac{2}{3})\;=\;
\frac{N_c\alpha_s}{\pi} 3 \ln 3 \; \approx 0.59
                                          \\
(d) & \,\, \chi (-\frac{1}{4}) \;=\;
\frac{N_c\alpha_s}{\pi} 6 \ln 2  \; \approx 0.75
\eeqn

We briefly discuss some properties of these results. Most striking
is the change in the power behavior in both $s$ and $M^2$.
Compared with both (4.17) or (4.18), the coefficients of $y_M$ and
$y_s$ have increased (cf.(4.27) and (4.28)): both values of the $\chi$-
function are larger than $\omega_{BFKL}=\chi(-\frac{1}{2}) =
\frac{4N_c \alpha_s}{\pi} 4 \ln 2 \approx 0.5)$. Translating this into
powers of $s$ and $M^2$, we obtain:
\beqn
(c) \,\, & \frac{d^2\sigma}{dtdM^2} \sim s^{2\chi(-\frac{1}{3})}
                      (M^2)^{-1-\chi(-\frac{1}{3})}  \\
(d) \,\, & \frac{d^2\sigma}{dtdM^2} \sim s^{2\chi(-\frac{1}{4})}
           (M^2)^{-1+\chi(-\frac{1}{4})-2\chi(-\frac{1}{2})} .
\eeqn
In contrast to the low-mass region (cases (a) and (b)), the momentum
distribution at the triple ladder vertex is now a result of diffusion
in $\ln(k_T^2)$ from both the upper and the two lower ladders. Combining
this with the conservation law $\mu=\mu_1+\mu_2$, we find that the
scale at the triple ladder vertex behaves as:
\beqn
(c)\,\, &\left< \frac{k_T^2}{Q_0^2}\right> \sim
       \left( \frac{Q^2}{Q_0^2} \right)^{1/3} \cdot
       \exp(-y_s \chi'(-\frac{1}{3}) )   \\
(d)\,\, & \left< \frac{k_T^2}{Q_0^2}\right> \sim
         \exp( -y_s \chi'(-\frac{1}{4})).
\eeqn
Again we find that the typical momentum scale at the triple ladder
tends to be smaller than $Q_0^2$. We finally mention that for
large $\ln(Q_0^2/Q_0^2)$ the momentum scale reaches, as the limiting
value, $Q_0^2$, i.e. it will never get large.

\subsection{The Four Gluon State}

So far our discussion had been restricted to the first part of the
triple Regge cross section (Fig.1b), the triple ladder vertex.
We now turn to the contributions
of Fig.1a. With our present understanding of the four gluon state
we are not yet able to calculate the cross section analytically: the
main obstacle is our ignorance of the leading $\omega$-plane singularity
of the four gluon state. Therefore we shall limit ourselves to
a qualitative discussion of the small-t behavior.
We first rearrange the interactions of the four gluon state as shown
in Fig.6a: first sum over all rungs in the channels (12) and (34),
then switch to the channels (13) and (24) etc. For the transition
vertex 2 gluons $\rightarrow$ 4 gluons which has been derived and
discussed in ~\cite{BW} we only need to know the following two
features: (i) it vanishes as any of the four lower gluon momenta
goes to zero; (ii) it possesses a scaling property:
when coupled to the BFKL ladders above and the two Pomerons below
(to be more precise: the upper $E^{(\nu)}$-functions of the two
BFKL-Pomerons), its dependence upon $q_1$ is simply a factor
$(q_1^2)^{\mu_1 + \mu_1 ' - \mu}$. The internal integrations
converge as long as $\mu > \mu_1 + \mu_2$, and for $\mu =
\mu_1 + \mu_2$ we get the familiar pole
\beqn
\frac{1}{\mu - \mu_1 - \mu_1'}
\eeqn
We have not yet attempted to calculate the coefficient, since our
dicussion of this part will have to remain qualitative anyhow.
For each ``switch`` from the
channels (ij), (kl) to (ik), (jl) or (il), (jk) (Fig.6b) we have an
effective (momentum dependent) vertex which scales as
\beqn
(q_i^2)^{-1 -\mu_i - \mu_i' + \mu_{i+1} + \mu_{i+1}'}
         W(q_{i+1}^2 / q_i^2).
\label{eff}
\eeqn
Let us say a few words about the function $W(q_{i+1}^2 / q_i^2)$.
In the ultraviolet region the internal momentum integral converges as
$dk^2/ (k^2)^2$. In the infrared region
a potential divergence could come from the region where one of the
internal lines (for example, line a in Fig.6b) becomes soft. As long as
the lower momenta ($q_{i+1}$ in Fig.6b) is nonzero, we have the
delta function pieces from both the Pomeron above and below which, at
first sight, seem to be ill-defined. However,
including the amputating factor $k^2$ and using the regularization
given in (\ref{gen}), we obtain the behavior
\beqn
 dk^2 \frac{ \lambda^2  }
           { (k^2)^{1- 2 \lambda }}
\eeqn
In the limit $\lambda \rightarrow 0$ we have two zeros in the numerator,
the singularity at $k^2=0$ is integrable, and the whole integral remains
finite.
When the lower momentum $q_{i+1}$ is taken to zero (i.e. in Fig.6b
the lower Pomerons are in the forward direction and behave as
$(k^2)^{-3/2-i\nu/2}$ (eq.(2.9)), both lines a and b become soft
simultaneously. Including the $\delta$-function pieces from the
upper two BFKL Pomerons, the $k$-integral diverges as $(q_{i+1}^2)
^{-1/2 -i\nu}$, but is multiplied by two $\lambda$-factors.
As a result, this divergent contribution drops out, and the
nonvanishing result comes from the nonsingular pieces of the
BFKL vertices. Hence $W$ is a well-defined function, and, in
particular, $W(0)$ is finite.
 From these simple arguments alone it already follows that the four
gluon state in Fig.6a
has the following pattern of momentum integrals (Fig6.c) :
\beqn
\int_0^{\infty} dq_1^2 ... dq_n^2 \frac{1}{\mu - \mu_1 - \mu_1'}
  (q_1^2)^{-1-\mu+ \mu_2 + \mu_2'} W(\frac{q_2^2}{q_1^2})
   \,\,   .... \,\,
  (q_n^2)^{-1 - \mu_n - \mu_n'+
                   \mu_l + \mu_r} W(\frac{q^2}{q_n^2})
\label{momi}
\eeqn
Introducing the new variables $\xi_1 = q_1^2 / q_2^2, \, ...,\,
\xi_n = q_n^2 / q^2$ we obtain:
\beqn
\int_0^{\infty} \frac{d \xi_1}{\xi_1} ... \frac{d \xi_n}{\xi_n}
      \frac{1}{\mu - \mu_1 - \mu_1'}
        \xi_1^{\mu_1 + \mu_1' - \mu} \,\,...\,\,
        \xi_n^{\mu_l + \mu_r - \mu}
        W(\frac{1}{\xi_1})...W(\frac{1}{\xi_n})
        (q^2)^{\mu_l + \mu_r - \mu}.
\eeqn
As long as $\mu>\mu_i + \mu_i'$ the $\xi$-integrals are finite, both
in the ultraviolet and infrared region. At $q^2=0$, the we put
$\xi_n = q_n^2 / Q_0^2$, and the integral over $\xi_n$ leads to
the conservation law $\delta(\mu - \mu_l - \mu_r)$.

In order to illustrate how the the saddle point argument works in
the limit $q^2 =0$ we restrict ourselves
to the case with only one two-Pomeron state above (Fig.7). To see the
interplay between the single BFKL and the two-BFKL state, it will be
useful to keep the rapidity variable at the $2 \rightarrow 4$ gluon
transition vertex. Instead of (\ref{4res}) we now have:
\beqn
\int_0^{y_M} dy_M'
\int \int \int \int \int \frac{d \mu}{2 \pi i}
     \frac{d \mu_1}{2 \pi i} \frac{d \mu_1'}{2 \pi i}
     \frac{d \mu_l}{2 \pi i} \frac{d \mu_r}{2 \pi i}
\frac{\left( \frac{q^2}{Q_0^2} \right)^{\mu_l + \mu_r - \mu}
      \left( \frac{Q^2}{Q_0^2} \right)^{\mu}       }
          {(\mu - \mu_1 - \mu_1')(\mu - \mu_l - \mu_r)}
              \nonumber \\
   \cdot     e^ {(y_M - y_M') \chi(\mu)
               + y_M'[\chi(\mu_1) + \chi(\mu_1')]
               + y_s [ \chi(\mu_l) + \chi(\mu_r)]}
\label{coup}
\eeqn
First, we consider $q^2$ to be of the order of $Q^2$. In this region
eq.(\ref{coup}) receives, from the nonforward coupling of the lower
Pomerons to the proton, an extra factor $\left( \frac{Q^2}{Q_0^2}
\right)^{-\mu_l - \mu_r}$, and the $Q^2$ dependence drops out.
As a result, the saddle point is solely determined by
the exponent $exp((y_M - y_M') \chi(\mu)
  + y_M'[\chi(\mu_1) + \chi(\mu_1')] + y_s [ \chi(\mu_l) + \chi(\mu_r)])$.
The saddle point analysis is simple and yields the value
$\mu_S={\mu_1}_S={\mu_1'}_S={\mu_l}_S={\mu_r}_S=-1/2$.
Inserting this value we find
$exp([y_M+y_M'+2y_s]\omega_{BFKL})$ which reaches its maximum at $y_M'=y_M$,
i.e. the two-BFKL state gets all the available rapidity. In the same
way the four gloun bound state will dominate over the single
BFKL-singularity which leads to a new power-behaviour in $M^2/s$.

If we take, now, $q^2$ to be of the order of $Q_0^2$, we have to perform a more
accurate saddle point analysis. Introducing $\mu_{l\;S}=\mu_{r\;S}
={\mu_{lr}}_S$ and $\mu_{1\;S}=\mu'_{1\;S}=\mu_S'$ we find the following
equations:
\be
0= - \chi(\mu_S)+ 2\chi(\mu_S')
\ee
\be
0=y_M' \chi(\mu_S')
\ee
\be
0=y_s \chi({\mu_{lr}}_S)
\ee
\be
0=(y_M-y_M') \chi'(\mu_S) + \ln(Q^2/q^2)
\ee
The second and third equations yield $\mu_S'={\mu_{lr}}_S=-1/2$.
The first equation then is solved by $\mu_S$ which
lies between -1/2 and -1. This saddle point gives the dominant behaviour
as long as $y_M$ exceeds the ``critical'' value
\beqn
y_{M \, c} = \frac{ \ln(Q^2/Q_0^2)}{|\chi'(\mu_S)|}.
\eeqn
For $y_M'$ we then find:
\be
y_M'= y_M - y_{M \, c}.
\ee
If $y_M$ is smaller than $y_{M \, c}$, $y_M'$ stays at zero, i.e         the
two-BFKL state has zero rapidity and becomes a subleading correction
to the single BFKL ladder above the two-BFKL state. Hence we are back
to the three-ladder case discussed before. In particular, if $y_M$
gets small in comparison with $\ln{Q^2/Q_0^2}$, the saddle point value
$\mu_S$ slides down towards $-1$, and the usual GLAP-dynamics takes
over.

In the limit $t\rightarrow 0$ we return to (4.38) and perform
the $\mu$-integral. Closing the contour to the left, we have the two
poles:
\beqn
\int_0^{y_M} dy_M'
\int \int \int \int
     \frac{d \mu_1}{2 \pi i} \frac{d \mu_1'}{2 \pi i}
     \frac{d \mu_l}{2 \pi i} \frac{d \mu_r}{2 \pi i}
     \frac{
 e^{y_M'[\chi(\mu_1) + \chi(\mu_1')] + y_s [\chi(\mu_l) + \chi(\mu_r)]}}
             {\mu_l + \mu_r - \mu_1 - \mu_1'}
   \nonumber  \\
 \left( \left( \frac{Q^2}{Q_0^2} \right)^{\mu_l + \mu_r}
        e^{(y_M - y_M')\chi(\mu_l+\mu_r)}
   -    \left( \frac{Q^2}{Q_0^2} \right)^{\mu_1 + \mu_1'}
    \left( \frac{q^2}{Q_0^2} \right)^{\mu_l + \mu_r - \mu_1 - \mu_1'}
        e^{(y_M - y_M')\chi(\mu_1+ \mu_1')}
          \right)
\label{coup2}
\eeqn
We will now show that, at $t=0$, only the first term survives;
to this end we study the saddle points of each term seperately.
We begin with the first one. Putting again $\mu_{l\;S}=\mu_{r\;S}
=\mu_{lr \,S}$, $\mu_{1\;S}=\mu'_{1\;S}=\mu_S'$, the saddle point
conditions are:
\be
0= - \chi(2 \mu_{lr \,S})+ 2\chi(\mu_S')
\ee
\be
0=y_M' \chi(\mu_S')
\ee
\be
0=(y_M-y_M') \chi'(2\mu_{lr \,S})
              + y_s \chi'(\mu_{lr \,S}) + \ln(Q^2/Q_0^2)
\ee
The second equation is solved if $\mu_S'=-1/2$ (the other possibility
$y_M'=0$ will be discussed in a moment). The first equation then is
solved by a value $\mu_{lr \,S}$ which lies between -1/2 and -1/3. As
long as $y_M$
exceeds (``large-M region``) the value
\beqn
y_{M \, c} = \frac{y_s \chi'(\mu_{lr \,S}) + \ln(Q^2/Q_0^2)}
                  {|\chi'(2 \mu_{lr \,S})|},
\eeqn
eq.(4.48) has the solution
\beqn
y_M'= \frac{y_M |\chi'(2 \mu_{lr \,S})| - y_S \chi'(\mu_{lr \,S})
                      -\ln(Q^2/Q_0^2) }
           {|\chi'(2 \mu_{lr \,S})| }
\eeqn
In other words, the rapidity $y_M$ is distributed between
the single BFKL Pomeron and the two-Pomeron state in a very
characteristic way. To obtain a result for the first part of
(4.45) we put $y_M'$ equal to the value given in (4.50) and
use the saddle point approximation for the $\mu$-integrals.
The result is proportional to:
\beqn
%\frac{2\pi}{(y_M-y_M')\chi''(2\mu_S) + y_s \chi''(\mu_S)}
%\frac{2\pi}{y_M' \chi''(-1/2)}
     \exp\left[2y_s \chi(\mu_{lr \,S}) + 2y_M \omega_{BFKL} +
                                    2 \mu_{lr \,S}\ln(Q^2/Q_0^2)\right]
\eeqn
Note that the coefficient $y_s$ is
 bigger than
the ``naive`` expectation $2 \omega_{BFKL}$.
When $y_M$ is lowered and reaches $y_{M \, c}$, the saddle point
value $y_{M\,S}$ moves down to zero, i.e. the single BFKL Pomeron state
gets the full available rapidity $y_M$. Now we are in a situation
analogous to the triple ladder approximation of the previous
subsection: the two-BFKL state becomes a subleading correction to the
triple ladder vertex. When $y_M$ further decreases
(or, alternatively, either $Q^2$ or $y_s$ become large), eqs.(4.46) and
(4.48) can no longer be satisfied: the maximum of the exponent in (4.45)
stays at $y_M'=0$, and $\mu_{lr \,S}$ slides down towards $-1/2$:
\beqn
\begin{array}{lclcl}
\mu_{lr \,S}&=& - \frac{1}{2}
             + \frac{1}{2}
\left( \frac{2y_M \alpha_s N_c}{y_s \pi \chi''(-1/2)} \right)
                   ^{\frac{1}{3} } & if &
 \ln(Q^2/Q_0^2) \ll  \left( y_M y_s^2 \chi'(-\frac{1}{2})^2
           \frac{\alpha_s N_c}{4 \pi} \right)^{\frac{1}{3}}
\\
\mu_{lr \,S} &=& - \frac{1}{2}
             + \frac{1}{2} \sqrt{\frac{y_M \alpha_s N_c}
                                      {\pi \ln(Q^2/Q_0^2)}}& if &
  \left( y_M y_s^2 \chi'(-\frac{1}{2})^2
           \frac{\alpha_s N_c}{4 \pi} \right)^{\frac{1}{3}}
                                         \ll \ln(Q^2/Q_0^2)
\end{array}
\eeqn
which is identical to the triple ladder case (4.15) and (4.16).

As to the second term in (4.45), the analysis is very similar
to the previous ones, and we can be brief in describing the main results.
The saddle point value $\mu_{lr \,S}$ follows from the condition
\be
0=y_s \chi'(\mu_{lr S}) - \ln(Q_0^2/q^2),
\ee
i.e. it starts at -1/2 and tends towards zero, as $q^2$ becomes smaller
and smaller. Again, there is a ``critical`` value of $y_M$:
\be
y_{M\;c} = \frac{\ln(Q^2/q^2)}
                {|\chi'(2 \mu_S')|}
\ee
where $\mu_S'$ satisfies
\be
\chi(2 \mu_S') = 2 \chi(\mu_S').
\ee
For $y_M > y_{M\;c}$ (or, alternatively, moderate $Q^2$ and finite
$q^2$), the maximal contribution belongs to some $0 < y_M' < y_M$,
i.e. the rapidity spreads over both the single and double
BFKL state. The power of $q^2$ is negative:
\be
\mu_l + \mu_r - \mu_1 - \mu_1' \approx 2 \mu_{lr \,S}- 2 \mu_S' <0
\ee
and the overall sign of the second term in (4.45) is positive.
For $y_M < y_{M\;c}$ (or, alternatively, smaller and smaller $q^2$), the
maximum comes from $y_M'=0$; the saddle point $\mu_S$ is now close
to zero, and $\mu_S'$ moves towards -1/2. As a result the exponent of
$q^2$ approaches +1, and the term vanishes $\sim q^2$.

Finally we wish to say a few words about the general case where
the iteration of the two-Pomeron states above the vertex generates a
new singularity in the $\omega$-plane. The structure of (\ref{momi})
shows that the conservation of conformal spin at $t=0$ works for
any iteration of the two-BFKL cut. Guided by the calculation of the
anomalous dimension of the four-gluon operator ~\cite{B,LRS} one may
speculate that the new singularity lies to the right of the two-BFKL
cut, i.e. $\omega_4 > 2\omega_{BFKL}$. As long as the corresponding
saddle point value of $\mu$ is different from -1 (we expect it again
to be at $\mu=-1/2$), the coupling of this
new singularity to the lower BFKL-singularities will have the same
features as in the triple ladder case, and we expect (qualitatively)
the same physical picture. To be concrete we expect an expression
similar to eq.(4.2), with
$\chi(\mu)$ being replaced by another, so far unknown function of
$\mu$. For the low mass region, the dominant behavior near $t=0$ is
given by the point $\mu=-1$, where GLAP-like evolution holds.
Consequently, the four gluon state will provide only
some (not so interesting) corrections to the triple ladder picture
(the higher-twist part of the four gluon state belongs to $\mu=-2$ and
is not important for our discussion here). In the large-$M^2$ region
the saddle point in $\mu$ will move away from -1 to some finite value
between -1 and 0, in analogy with (4.23), (4.24). Because of the higher
intercept, the four gluon state will obtain the full rapidity $y_M$,
whereas the single-BFKL state acts like a direct coupling of the
four gluon system to the fermion loop. Furthermore, in this region of
$y_M$ the contribution of the four gluon state will dominate over the
triple ladder part.

\subsection{Summary}

Let us try to summarize the results obtained in this section.
Starting at some value away from $t=0$, say $q^2=-t$ of the order of the
hadronic scale $Q^2$, our cross section formula will be dominated by
the leading $\omega$-plane singularities above and and below the
triple vertex. In the upper t-channel the leading singularity is
given by the four gluon state - either the two-BFKL state or a new
four gluon bound state with intercept $\omega_4 > 2 \omega_{BFKL}$ -,
whereas in the lower Pomerons we have the usual BFKL singularity at
$\omega=\omega_{BFKL}$. In this region of momentum transfer $t$
one observes a negative power of $t$, i.e. the cross section grows
with decreasing $-t$ as $1/\sqrt{-t}$.
As to the momentum scale at the triple vertex,
we have the usual diffusion picture in all three t-channels. In
all three Pomerons, the diffusion into the infrared region is stopped
by the momentum transfer $t\sim Q^2$, only a small
contribution might come from the region at or below the hadronic scale $Q_0^2$
unless the photon mass $Q^2$ is to small (of the order of $Q_0^2$).

As $t$ approaches zero, several changes occur. First of all, the
negative power of $t$ starts to move towards positive values: the
initial rise with $t$ comes to stop, and the cross section reaches a
finite limit. At $t=0$, the dependence upon $s$ and $M^2$ becomes
rather involved. Most striking, there is no simple uncorrelated energy
dependence, but, the powers of $s$ and $M^2$ change with the kinematic region.
To begin with the small-M region (the precise condition is given in
(4.44)), the upper gluon system is determined by GLAP dynamics,
i.e. we have a clean twist-two state with strong ordering in the
transverse momentum. The lower Pomerons are to be evaluated in the
BFKL limit and the singularity at $\omega_{BFKL}$ desribes the
$s$ dependence. In particular, the four-gluon state above the triple
vertex is nonleading: it serves merely as a renormalization of the
triple ladder vertex discussed in the first part. The typical
momentum scale at the triple ladder vertex arises from the competition
between the strong ordering dynamics above and the diffusion mechanism
from below: the former one tends to push the average scale into
the infrared region, even below the hadronic scale $Q_0^2$.

The large-$M$ region, on the other hand, has quite different
characteristics. Generally speaking, now the four gluon state above
the triple vertex is equally or even more important than the simple
triple ladder vertex. We therefore have to consider the
contribution of the two-BFKL singularity and, in case it exists, also
the formation of a new bound state to the right of $2\omega_{BFKL}$.
In more detail, due to the conservation in $\mu$ (conformal dimension),
the leading $\omega$ singularities in the upper and the
lower Pomerons are linked together which
implies, for the lower Pomerons, that the leading $\omega$ is
larger than $\omega_{BFKL}$. The amount by which the singularities
are shifted towards larger
values, depends upon the way in which the total available rapidity
is distributed between the missing mass ($y_M = \ln(M^2/Q^2)$)
and the rapidity gap ($y_s=\ln(s/M^2)$).
One of the main conclusions
of this analysis of the energy dependence therefore is that
{\it the way in which the BFKL Pomeron contributes depends upon its
environment.}\\

\section{Discussion and Conclusions}
\setcounter{equation}{0}

In this paper we have obtained first analytic results on
the rather complicated cross section formula for the diffractive
dissociation of the photon in deep inelastic scattering of
{}~\cite{BW}. Our main interest was the behaviour near $t=0$:
for several reasons we expect this point to be particularly
``dangerous`` for the validity of perturbation theory.
As one of the main results of our investigation we have found
that, within the BFKL approximation, the cross section is finite at
zero momentum transfer $t=-q^2=0$. At the same time, however, the
$\ln{k_t^2}$ diffusion has penetrated deeply into the infrared region,
and the typical transverse momentum at the triple Pomeron vertex is
fairly small. Consequently, the BFKL approximation used in ~\cite{BW}
provides a well-defined starting point for a sytematic unitarization
procedure, but it also emphazises the need for including higher order
unitarizing corrections. Performing a careful saddle point analysis of
our cross section formula we have also found very special features
of the dependence upon $s$ and $M^2$ near $t=0$. As a function of $t$,
the cross section has a cusp-structure, and the cusp shrinks with
increasing energy. The origin of these phenomena can be traced back to
the conservation of conformal dimensions which relates to one of the
profound properties of the BFKL approximation.

Although our main interest concerns the limit $t=0$, it is instructive
to extend our discussion to the region of nonzero $t$. We consider the
triple Regge
limit with $\ln(Q^2/Q_0^2)$ being of the order of $\ln(M^2/Q^2)$.
Starting with $t$ of the order of $Q^2$, we find that the upper part of the
diagram 1a,1b is governed by the usual diffusion of the
internal transverse momenta around $Q^2$. This kinematic region leads
to the $1/\sqrt{-t}$-behaviour of the cross section ~\cite{MueP},
and in the simplest case of only three ladders to the coupling of the
three BFKL-singularities. New in our analysis is the four gluon state
above the junction of the two lower Pomerons: it contains
the two-BFKL state, but we expect that also a new bound state to the
right of the two-BFKL singuilarity will be formed. Both the two-BFKL
state and the new bound state will dominate the single-BFKL state, and
we expect a different, new dependence on $M^2$ which may be measured in
future.

Decreasing $-t$ down to $Q_0^2$ we have to distinguish between two
different cases: the ``high mass region`` ($\ln{M^2/Q^2}$ larger
than $\ln{Q^2/Q_0^2}$) and the ``low mass region`` ($\ln{M^2/Q^2}$
smaller than $\ln{Q^2/Q_0^2}$); the more precise definition is given in
section 4 (eq.(4.43)). In the former case we are still in diffusion
region, and the four gluon state plays an important role. In the
second case, however, the case of not so large $M^2$, we enter the GLAP
region where the single BFKL ladder gives a larger contribution than
the two-BFKL state or the new bound state since both contributions are
subleading in $\ln(Q^2/Q_0^2)$. Now the dynamic of the evolution has
changed crucially from diffusion (which includes all higher twist
contributions) to the usual GLAP-evolution and the dominance of the
leading twist piece. Furthermore, the $M^2$-dependence experiences a
change which should be measurable.

The advantage of considering first the region $t \neq 0$ (before moving
towards $t=0$) is that the BFKL diffusion stays away from the
infrared region, and the use of perturbation theory is better
justified. From this point of view it would be even more advantageous
to move into the large
t-region, $-t\gg Q^2$. A kinematic configuration of this type is
realized in the diffractive vector meson production as discussed in
ref. \cite{FR} (see also ref.\cite{BFLLRW}). The approach in ref.
\cite{BW} is general enough to be also applicable for this process.
One only needs to substitute the corresponding wave
functions of the initial particles and has to carry out an analysis
quite analogous to the one outlined in this paper. The major difference
becomes visible when trying to perform the saddle point analysis:
for $q^2= -t \rightarrow \infty$ (as opposed to: $q^2 \rightarrow 0$)
the $\mu$ contours have to be closed in the opposite direction, and
other singularities become relevant. The large-$t$ limit, therefore,
requires a separate investigation.

Returning to $t=0$ we again have the two regions of low and high mass
$M$ (the precise definition is in (4.49) and now also depends upon
$\ln( s/M^2 )$). As before, the high mass region is characterized by
the diffusion dynamics, and the four gluon state plays an important
role, whereas the low mass region is governed by the single BFKL ladder
in the GLAP region. At the same time, however, we face the problem that
the typical momentum scale at the Triple Pomeorn vertex lies far in the
infrared region where leading order perturbation theory becomes
unreliable and we should compute higher order correction.
At the moment, therefore, we can only guess what the ``true`` QCD
behavior will be. Here a
comparison of the small-x behavior of $F_2$ at large $Q^2$ with the
high energy behavior of the photoproduction total cross section may
be helpful. For sufficiently large $Q^2$, it seems quite adequate to
use the BFKL-approximation for the x-dependence of $F_2$, since the
main contribution of the momentum integrals comes from the
ultraviolet region; the result of this is the well-known power
behavior $F_2 \sim (1/x)^{\omega_{BFKL}}$. When we lower $Q^2$, the
contribution of small internal momenta becomes larger and we should
include more and more corrections to the BFKL-approximation;
eventually, nonperturbative effects will take over.
On the other hand we know that at $Q^2=0$ (the photoproduction limit)
the energy dependence is much weaker ($\sigma_{tot}(\gamma^*) \sim
(W^2)^{0.08}$) than at large $Q^2$:
as a first guess, therefore, one expects that the ``true`` (as
opposed to: perturbative) QCD-behavior in the infrared region will
tend to lower the increase with the energy. Consequently, in our
cross section formula (4.1) for the diffractive dissociation we expect
that at $t=0$ the effective power of $s/M^2$ will be smaller than
predicted by our perturbative analysis. How this combines with the
$t$-dependence obtained in our analytic analysis has to be studied in
a numerical analysis which will be the next step in our program.

A last remark should be made on the Pomeron structure function.
Although the terminology 'Pomeron structure function' is questionable
since factorization does not hold in the usual sence, we will
nevertheless use it here, since in the literature diffractive
dissociation is fairly often discussed in those terms. One of the new
elements included into our analysis is the four gluon state in the
upper t-channel. Since the two lower Pomerons already come with
two gluons each, this four gluon state comes for free, i.e. it
costs now extra power of $\alpha_s$ to create this state. In this sense
the appearance of the four gluon state in the Pomeron structure
function is not as much a higher order effect as in $F_2$.
One also should bear in mind that, within the BFKL approximation,
the four gluon state not only contributes to twist four but also to
the leading twist. As to the question under which circumstances this
four gluon state contributes, we distinguish between the two cases
mentioned before (low mass and high mass). In the former case
we have GLAP-dynamics above the triple Pomeron vertex, i.e. the $Q^2$
evolution is desribed by the usual (leading twist) evolution equations.
The four gluon state only appears in the initial distribution.
For the latter case, the upper part is governed by BFKL-type diffusion,
and we have seen that the four gluon state may even dominate the
two-gluon ladders. Here the $Q^2$- evolution will feel the presence of
the four gluon state (to describe this in more detail requires a better
understanding of the dynamics of the four gluon state). In summary,
in the large-mass region we expect the Pomeron structure function to be
more effected by the new four gluon state (``screening'') than
$F_2$, the total deep inelastic cross section.

{\bf Acknowledgements:} We gratefully acknowledge valuable discussions
with M.Ryskin. One of us (J.B.) wishes to thank Al Mueller
for very useful conversations.

\vspace{1cm}
\section*{Figure captions}
\begin{description}
\item Fig. 1a  :
                General structure of the cross section of  the
                diffractive dissociation of a virtual photon into
                $q\bar{q} + n$ gluons in the triple regge limit.
                Here and in the following figures
                wavy lines denote reggeized gluons,
                                                  a shaded
                circle represents the BFKL-pomeron
                (eq.\ref{pom}), a shaded ellipse represents
                the pomerons coupling to the proton
                (eq.\ref{had}) and black ellipses represent
                different types of $2 \rightarrow n$
                gluon interactions.
\item Fig. 1b  :
                A disconnected contribution to the cross of
                $\gamma^{\ast}+p \rightarrow p +   q\bar{q} +
                 n $ gluons.
\item Fig. 2a  :
                Amplitude of the production of a $q\bar{q}$ -
                pair in deep inelastic diffractive photon-proton
                scattering.
                The summation runs over all different couplings
                of the two reggeized gluons to the
                $q\bar{q}$ - pair.
\item Fig. 2b  :
                Graphical representation of the building blocks
                of the partial wave amplitude (eq.\ref{pwf})
                of diffractive $q\bar{q}$ - production.
\item Fig. 3   :
                Graphical representation of the total
                cross section $\sigma(\gamma^{\ast} + $ proton)
                (eq.(\ref{tot}).
\item Fig. 4   :
                A contribution to the amplitude of production
                of $q\bar{q} + n$ gluons in
                        deep inelastic diffractive photon-proton
                scattering.
\item Fig. 5   :
                Integration path and singularity
                of the $\mu$ - integration (eq.\ref{4res})
                in the
                complex $\mu$ - plane.
\item Fig. 6a  :
                Reordering of two-gluon interactions in the
                four gluon state.
\item Fig. 6b  :
                Graphical representation of the effective
                vertex (eq.\ref{eff})
                for the recoupling from interaction channels
                (ij),(kl) to (ik),(jl).
\item Fig. 6c  :
                Compact representation of the four gluon
                state as a state of two pomerons interacting
                via an effective recoupling vertex.
\item Fig. 7   :
                 The contribution (\ref{coup})
                 to the cross section of diffractive production
                 of $q\bar{q} +$ gluons with a single pomeron
                 interaction in the four gluon state.
\end{description}
\newpage
\setlength{\unitlength}{1cm}
\setcounter{figure}{1}
\alphfig
\begin{figure}[t]
\begin{center}
\begin{picture}(8,8)(0,0)
%\put(0,0){\framebox(8,8)}
\epsfig{file=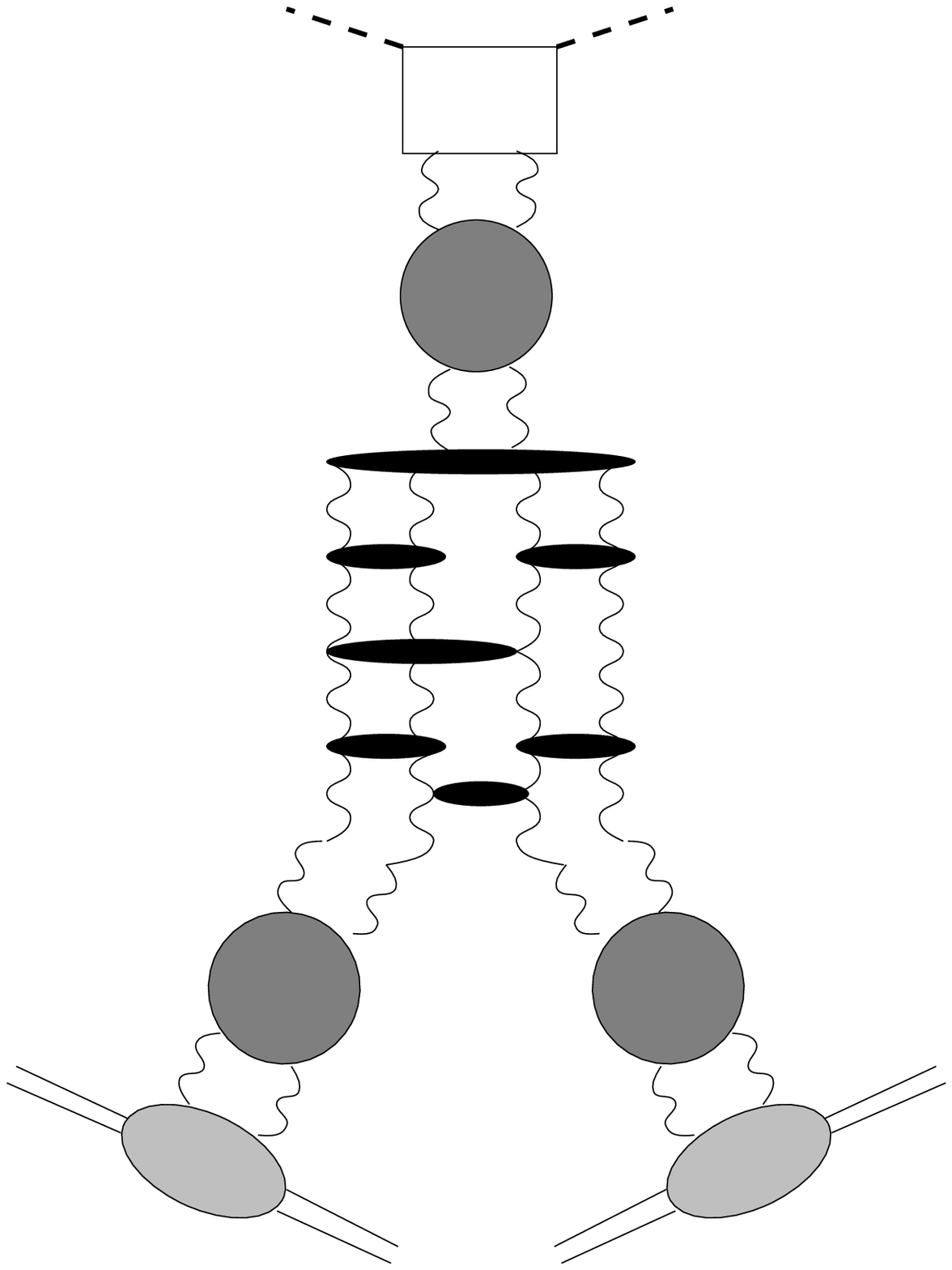,height=9.6cm,width=8cm}
\end{picture}
\caption{}
\label{fig1a}
\end{center}
\end{figure}
\begin{figure}[b]
\begin{center}
\begin{picture}(8,8)(0,0)
%\put(0,0){\framebox(8,8)}
\epsfig{file=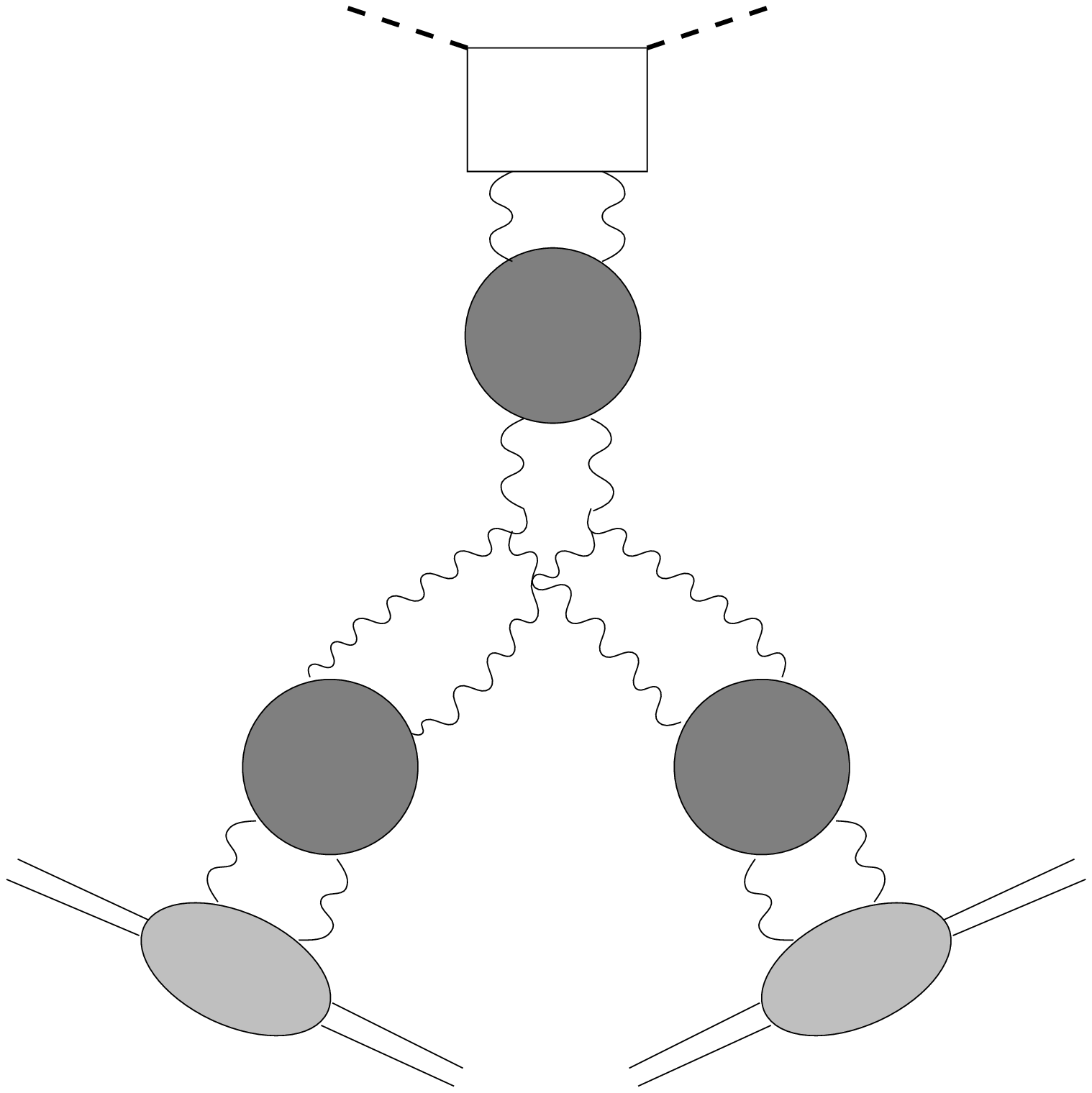,height=8cm,width=8cm}
\end{picture}
\caption{}
\label{fig1b}
\end{center}
\end{figure}
\setcounter{figure}{2}
\alphfig
\begin{figure}[t]
\begin{center}
\begin{picture}(8,8)(0,0)
%\put(0,0){\framebox(8,8)}
\epsfig{file=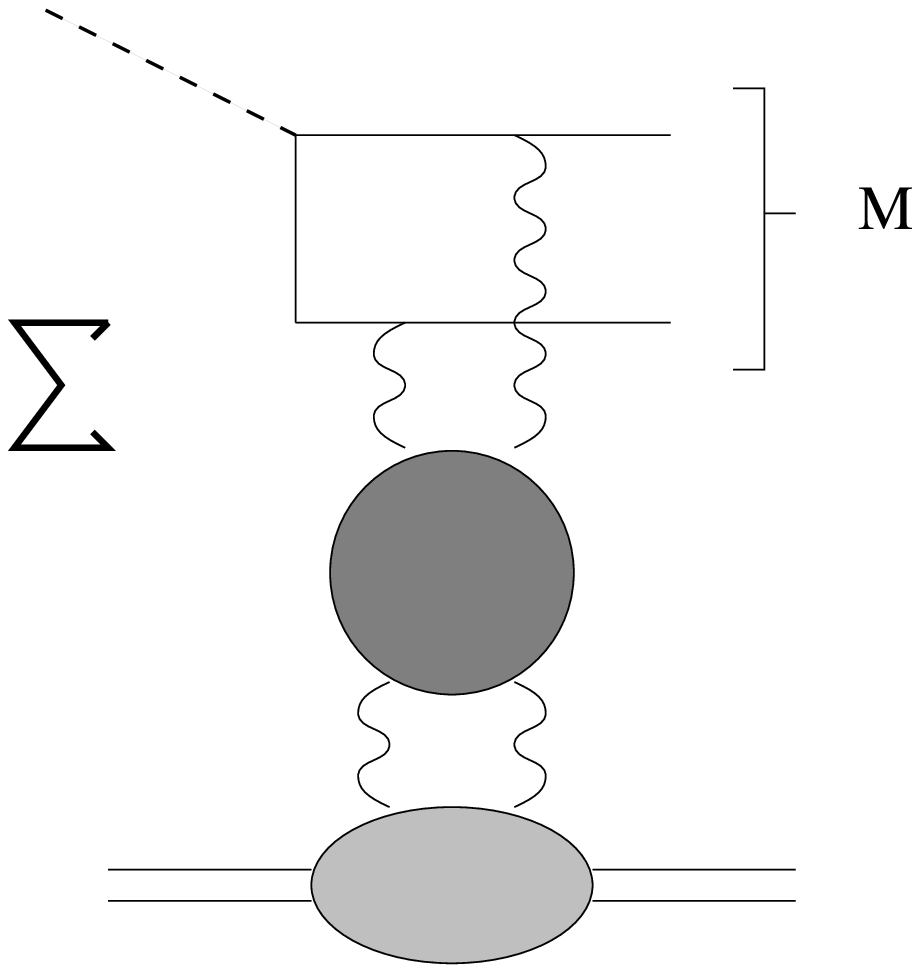,height=8cm,width=8cm}
\end{picture}
\caption{}
\label{fig2a}
\end{center}
\end{figure}
\begin{figure}[b]
\begin{center}
\begin{picture}(8,8)(0,0)
%\put(0,0){\framebox(8,8)}
\put(-3.1,0){\epsfig{file=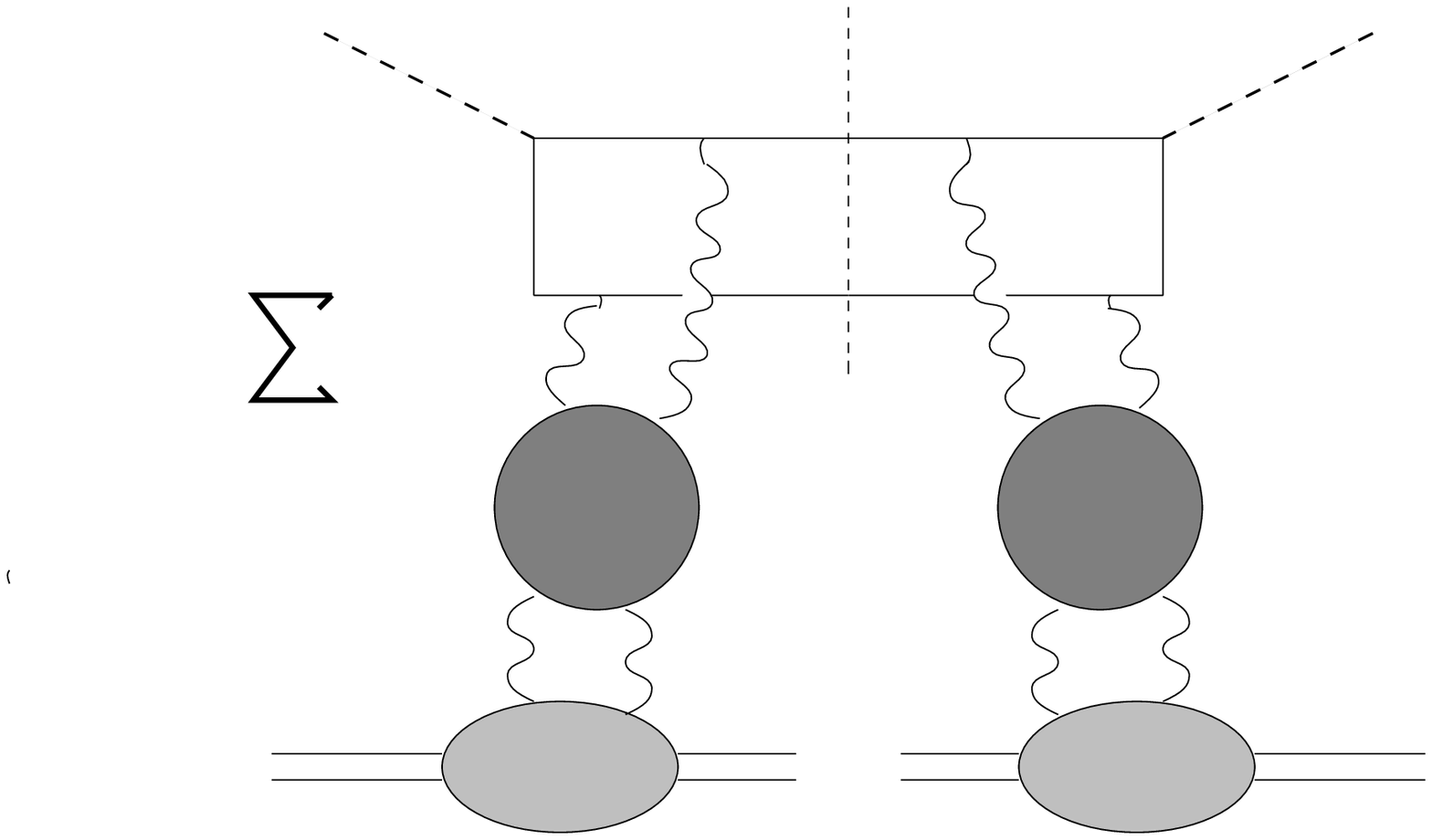,height=8cm,width=12cm}}
\end{picture}
\caption{}
\label{fig2b}
\end{center}
\end{figure}
\resetfig
\begin{figure}[t]
\begin{center}
\begin{picture}(8,8)(0,0)
%\put(0,0){\framebox(8,8)}
\put(0,0){\epsfig{file=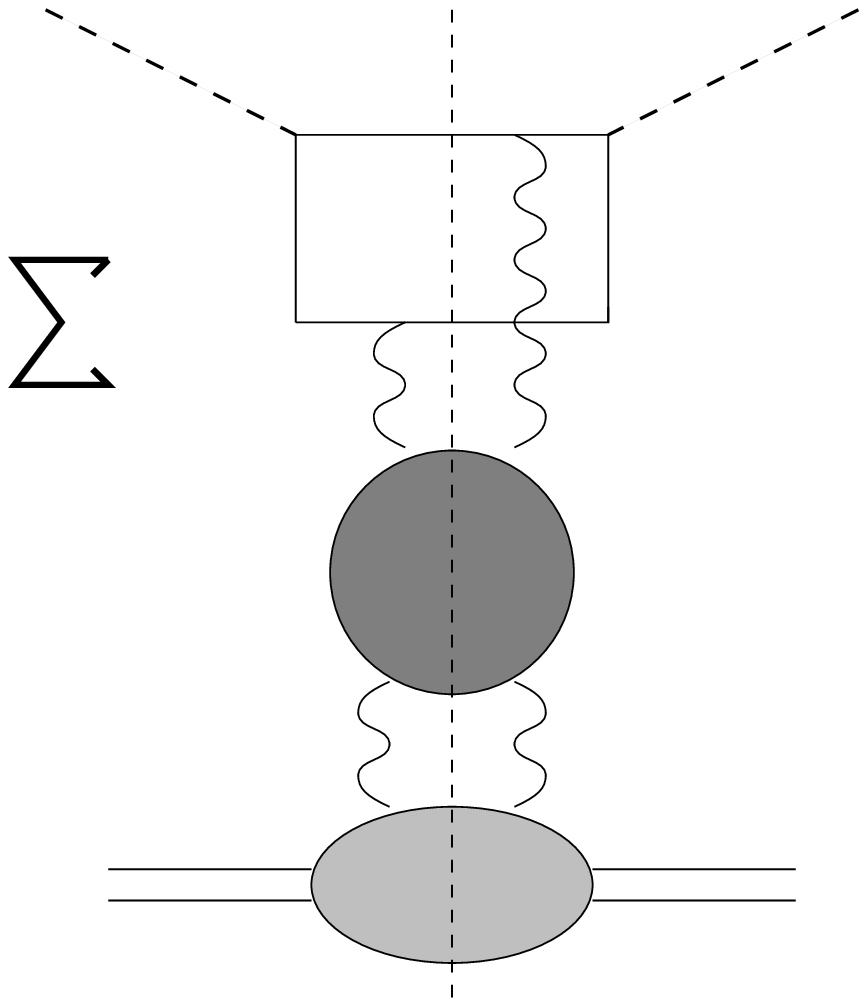,height=8cm,width=8cm}}
\end{picture}
\caption{}
\label{fig3}
\end{center}
\end{figure}
\begin{figure}[b]
\begin{center}
\begin{picture}(8,8)(0,0)
%\put(0,0){\framebox(8,8)}
\put(1.1,0){\epsfig{file=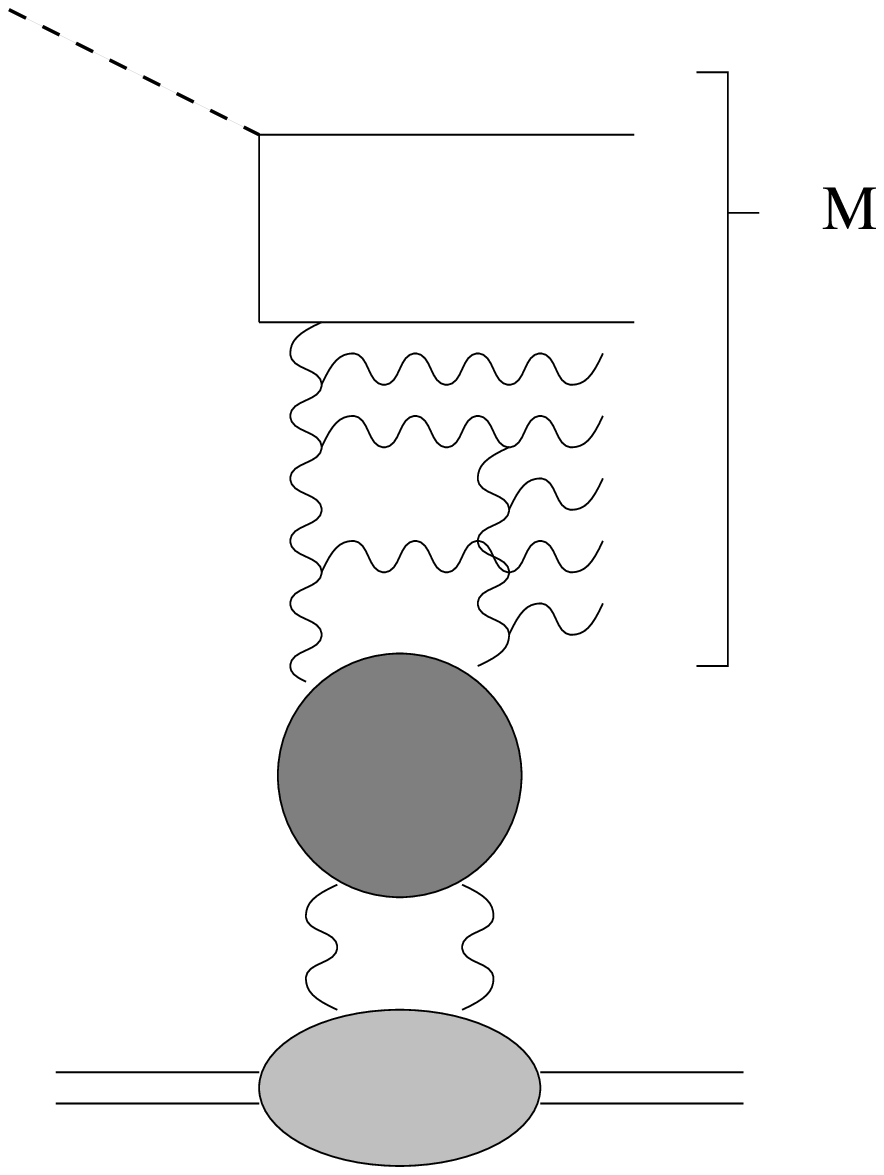,height=8cm,width=6.4cm}}
\end{picture}
\caption{}
\label{fig4}
\end{center}
\end{figure}
\begin{figure}[t]
\begin{center}
\begin{picture}(8,8)(0,0)
%\put(0,0){\framebox(8,8)}
\epsfig{file=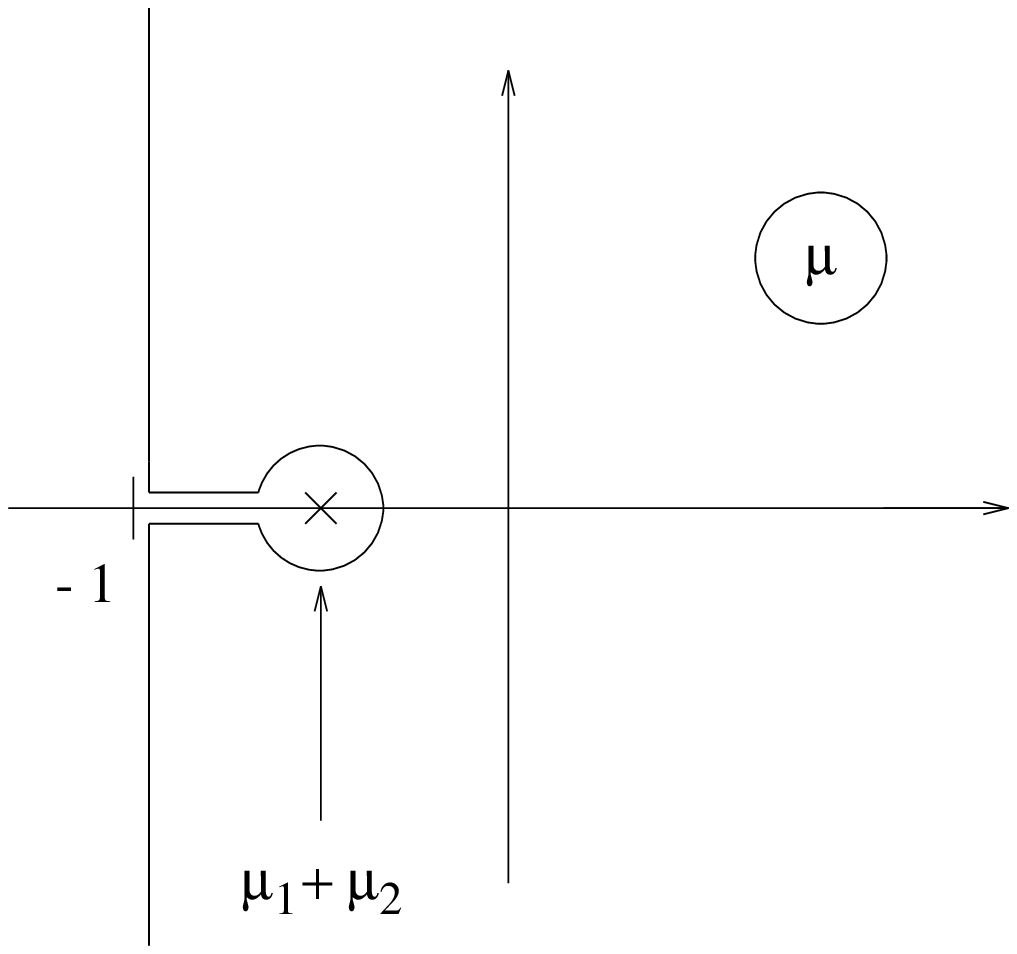,height=8cm,width=8cm}
\end{picture}
\caption{}
\label{fig5}
\end{center}
\end{figure}
\setcounter{figure}{6}
\alphfig
\begin{figure}[t]
\begin{center}
\begin{picture}(8,8)(0,0)
%\put(0,0){\framebox(8,8)}
\put(-1.35,0){\epsfig{file=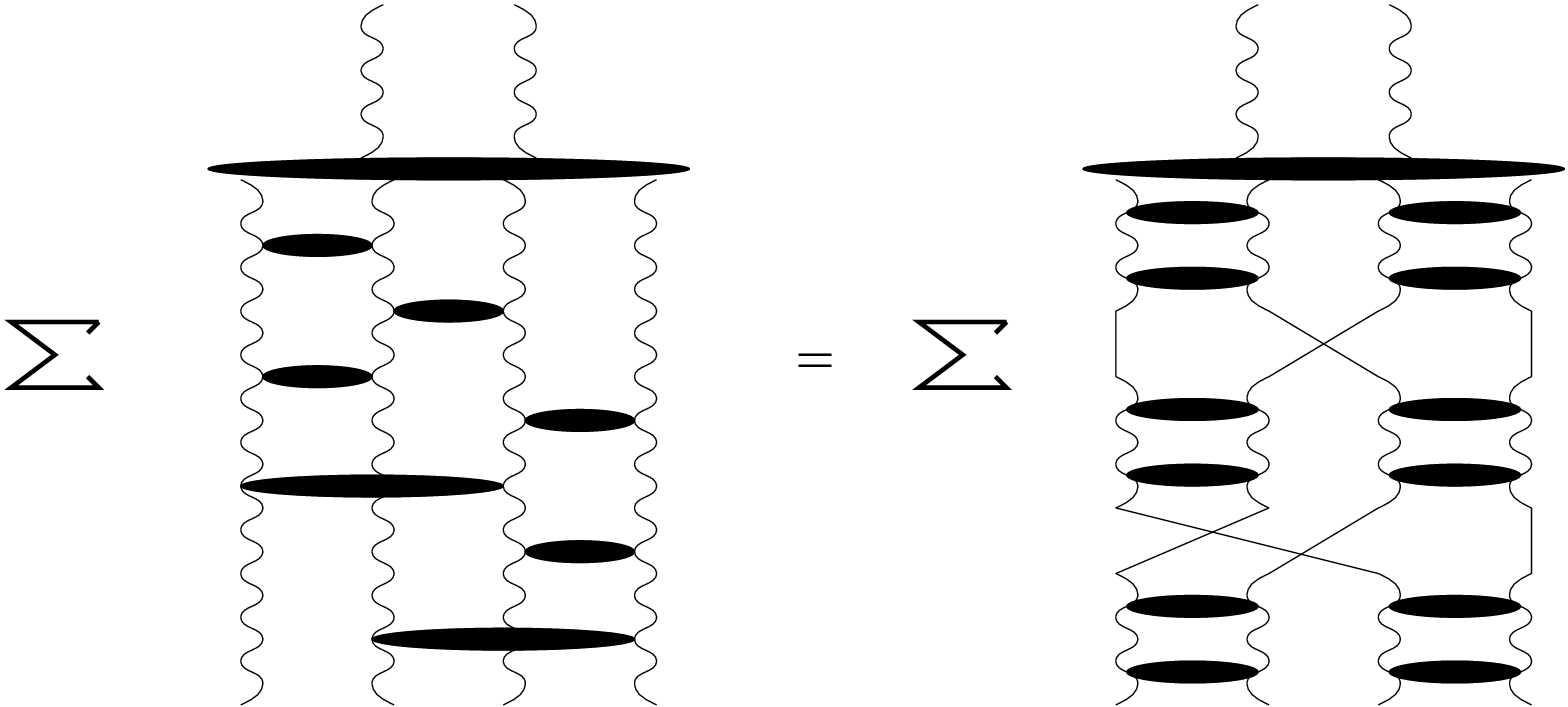,height=6.4cm,width=9.6cm}}
\end{picture}
\caption{}
\label{fig6a}
\end{center}
\end{figure}
\begin{figure}[t]
\begin{center}
\begin{picture}(8,8)(0,0)
%\put(0,0){\framebox(8,8)}
\put(2,2){\epsfig{file=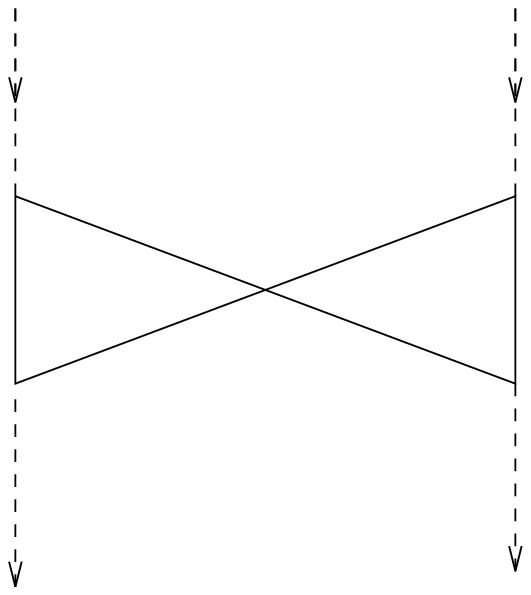,height=4cm,width=4cm}}
\put(1.5,1.5){${\bf q}_{i+1}$}
\put(6,1.5){$-{\bf q}_{i+1}$}
\put(1.5,6.2){${\bf q}_{i}$}
\put(6,6.2){$-{\bf q}_{i}$}
\put(2.2,2.2){$(ik)$}
\put(5.1,2.2){$(jl)$}
\put(2.2,5.5){$(ij)$}
\put(5.1,5.5){$(kl)$}
\put(1.3,3.85){(a)}
\put(2.7,3.3){(b)}
\end{picture}
\caption{}
\label{fig6b}
\end{center}
\end{figure}
\begin{figure}[b]
\begin{center}
\begin{picture}(8,8)(0,0)
%\put(0,0){\framebox(8,8)}
\epsfig{file=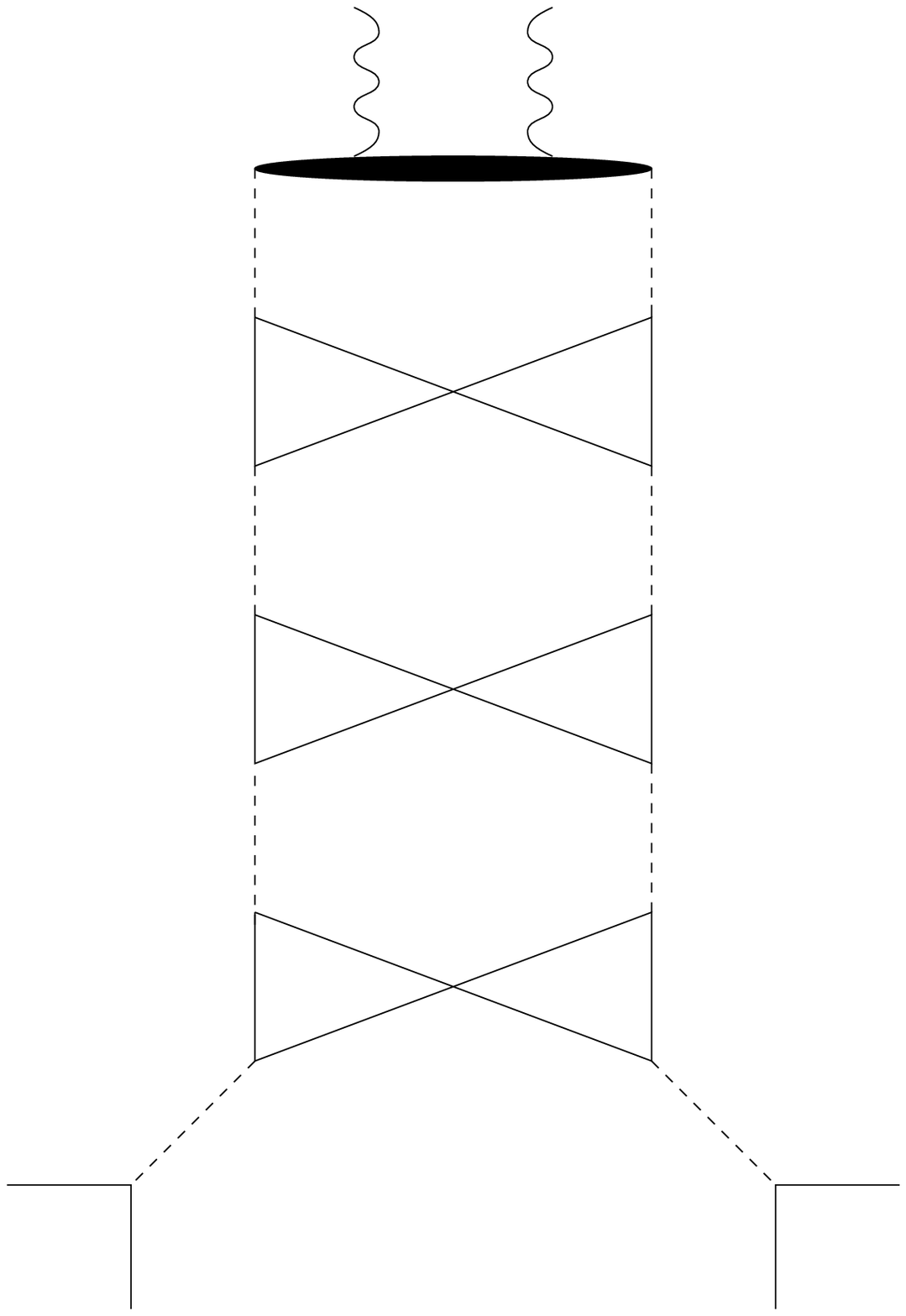,height=10cm,width=8cm}
\put(-5.5,8.1){${\bf q}_1$}
\put(-5.5,5.7){${\bf q}_2$}
\put(-5.5,3.3){${\bf q}_n$}
\put(-6,1.2){${\bf q}$}
\put(-6.5,8.1){$\mu_1$}
\put(-2.0,8.1){$\mu'_1$}
\put(-6.5,5.7){$\mu_2$}
\put(-2.0,5.7){$\mu'_2$}
\put(-6.5,3.3){$\mu_n$}
\put(-2.0,3.3){$\mu'_n$}
\put(-7,1.5){$\mu_l$}
\put(-1.5,1.5){$\mu_r$}
\put(-5.5,9.25){$\mu$}
\end{picture}
\caption{}
\label{fig6c}
\end{center}
\end{figure}
\resetfig
\setcounter{figure}{6}
\begin{figure}[t]
\begin{center}
\begin{picture}(8,8)(0,0)
%\put(0,0){\framebox(8,8)}
\put(1.75,0){\epsfig{file=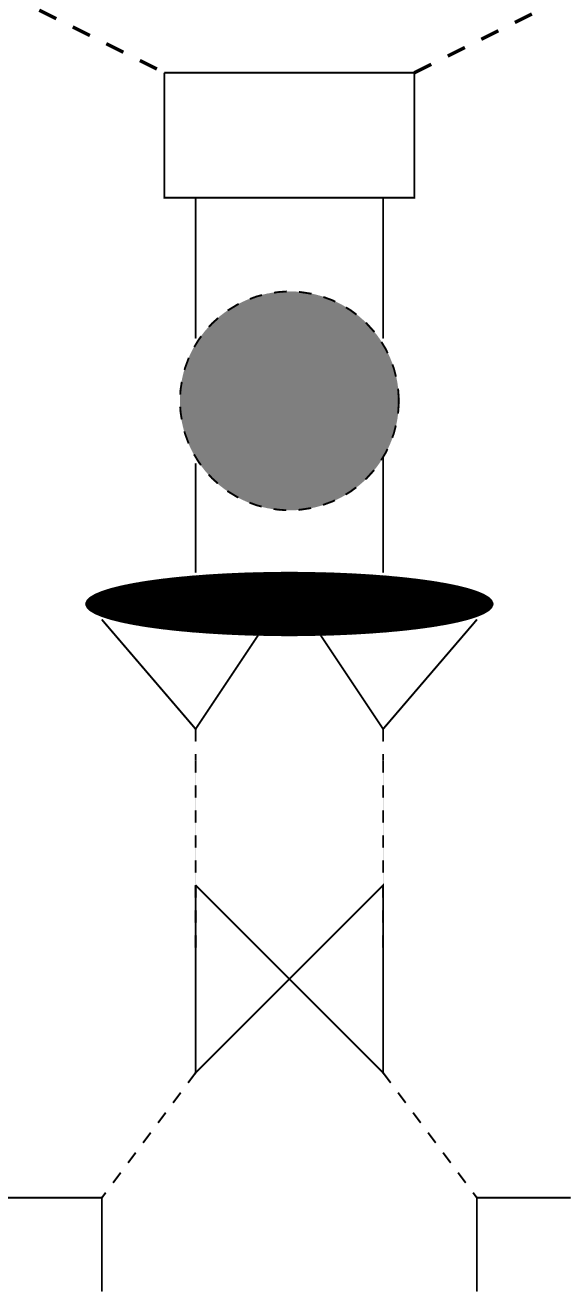,height=8cm,width=4.4cm}}
\put(2.3,1){$\mu_l$}
\put(5.5,1){$\mu_r$}
\put(2.6,2.8){$\mu_1$}
\put(5,2.8){$\mu'_1$}
\put(3,0.7){${\bf q}$}
\put(4.5,0.7){$-{\bf q}$}
\end{picture}
\caption{}
\label{fig7}
\end{center}
\end{figure}
\end{document}